\documentclass[12pt]{article}
\pdfoutput=1
\usepackage{jheppub}
\usepackage{verbatim}

\newcommand{\be}{\begin{equation}}
\newcommand{\ee}{\end{equation}}
\newcommand{\HH}{\mathcal{H}_H}
\newcommand{\HA}{\mathcal{H}_A}
\newcommand{\HB}{\mathcal{H}_B}
\newcommand{\tr}{\mathrm{tr}}
\newcommand{\HR}{\mathcal{H}_R}

\newcommand{\lan}{\langle}
\newcommand{\ran}{\rangle}

\newcommand{\mO}{\mathcal{O}}
\newcommand{\tO}{\widetilde{\mathcal{O}}}
\newcommand{\wt}{\widetilde}
\newcommand{\A}{\mathcal{A}}
\begin{document}
%\subheader{empty}
\title{Aspects of the Papadodimas-Raju Proposal for the Black Hole Interior}
\author[a]{Daniel Harlow}
\affiliation[a]{Princeton Center for Theoretical Science, Princeton University, Princeton NJ 08540 USA}

\emailAdd{dharlow@princeton.edu}
\abstract{In this note I elaborate on some features of a recent proposal of Papadodimas and Raju for a CFT description of the interior of a one-sided AdS black hole in a pure state.  I clarify the treatment of $1/N$ corrections, and explain how the proposal is able to avoid some of the pitfalls that have disrupted other recent ideas.  I argue however that the proposal has the uncomfortable property that states in the CFT Hilbert space do not have definite physical interpretations, unlike in ordinary quantum mechanics.  I also contrast the ``state-dependence'' of the proposal with more familiar phenomena, arguing that, unlike in quantum mechanics, the measurement process (including the apparatus) in something like the PR proposal or its earlier manifestations cannot be described by unitary evolution.  These issues render the proposal somewhat ambiguous, and it seems new ideas would be needed to make some version of it work.  I close with some brief speculation on to what extent quantum mechanics should hold for the experience of an infalling observer.}
\maketitle

\section{Introduction}
The question of whether or not the physics of black holes is described by quantum mechanics has a long history, going back to the seminal papers of Hawking \cite{Hawking:1974sw,Hawking:1976ra}.  The majority of people working in the field now believe that it is, motivated primarily by the success of the BFSS matrix model \cite{Banks:1996vh} and especially the AdS/CFT correspondence \cite{Maldacena:1997re,Witten:1998qj,Gubser:1998bc}.  In these examples one has a fully quantum mechanical description of black hole formation and evaporation, so the issue of whether it is possible to have a theory that describes black holes quantum mechanically appeared to be settled.  There were always lingering doubts however about how exactly these quantum mechanical theories could be used to describe the experience of the infalling observer \cite{Mathur:2009hf,Giddings:2011ks}.  In the last few years this lingering uncertainty has been crystallized into a relatively sharp set of paradoxes, all of which seem to imply that a description of the infalling observer requires some sort of extension or modification of the quantum mechanical theory used to describe the formation and evaporation of the black hole \cite{Braunstein:2009my,Almheiri:2012rt,Marolf:2012xe,Almheiri:2013hfa,Bousso:2013wia,Marolf:2013dba}.  

 Recently Papadodimas and Raju have made an interesting proposal for the description of the black hole interior in AdS/CFT \cite{Papadodimas:2013wnh,Papadodimas:2013jku}.  Their proposal is related to earlier ideas that are often grouped together under the slogan ``$A=R_B$'' \cite{Bousso:2012as,Susskind:2012uw,Papadodimas:2012aq,Verlinde:2012cy}, or somewhat more carefully ``ER=EPR'' \cite{Maldacena:2013xja,Susskind:2013lpa},\footnote{I here mean some version of these ideas which would prevent firewalls in generic states.  Motivated by complexity-theoretic arguments Susskind has recently been exploring the possibility of a version where generic states would have firewalls, but black holes formed by short collapses would not \cite{Susskind:2014rva,Susskind:2014ira}.} but the new proposal is considerably more precise than any of this previous work.  It moreover is able to cleverly avoid some of the objections \cite{Almheiri:2013hfa,Marolf:2013dba,Bousso:2013ifa} raised to $A=R_B$ (or to ER=EPR).  The main new idea is to focus on a ``small algebra'' of potential observables, with respect to which one defines a set of ``equilibrium states'' that are expected to have smooth horizons.  For each operator in the small algebra one can then define a ``mirror operator'', which for the case of the mirror of an operator related to a mode just outside the black hole horizon has the interpretation of acting on a mode just behind the horizon, which has been difficult to get at by other means.  The most controversial part of the proposal, which it inherits from $A=R_B$ or ER=EPR, is that the mirror operators are defined in a ``state-dependent'' manner, which is not allowed in ordinary quantum mechanical measurement theory.

In this note I attempt a careful critical analysis of the PR proposal.  In section \ref{proposalsec} I will introduce the proposal, clarifying some aspects that were not completely transparent in the original papers, especially the treatment of $1/N$ corrections.  In section \ref{nonusec} I argue that in the PR proposal, pure quantum states in the CFT associated with large black holes do not have definite physical interpretations for the infalling observer, unlike in ordinary quantum mechanics.  In section \ref{genbhsec} I discuss new issues that arise in extending the proposal to more general black holes, namely two-sided AdS wormholes and evaporating Minkowski black holes.  For the AdS wormholes I consider several possible extensions of the one-sided proposal, identifying one which I consider to be the most appealing.  For evaporating black holes, there is a new problem in that the small algebra does not seem to be sufficient for describing the realm of possible experiments.  In section \ref{sdsec} I study the ``state-dependence'' of the proposal in more detail, emphasizing the considerable extent to which it violates quantum mechanics.  I compare this to more conventional physical situations where naively state dependent operators arise but the measurement theory is nonetheless consistent with quantum mechanics.  Finally I close with some brief remarks on the expected validity of quantum mechanics for the infalling observer.  The later sections can basically be read in any order.

\section{Description of the Proposal}\label{proposalsec}
\begin{figure}
\begin{center}
\includegraphics[height=7cm]{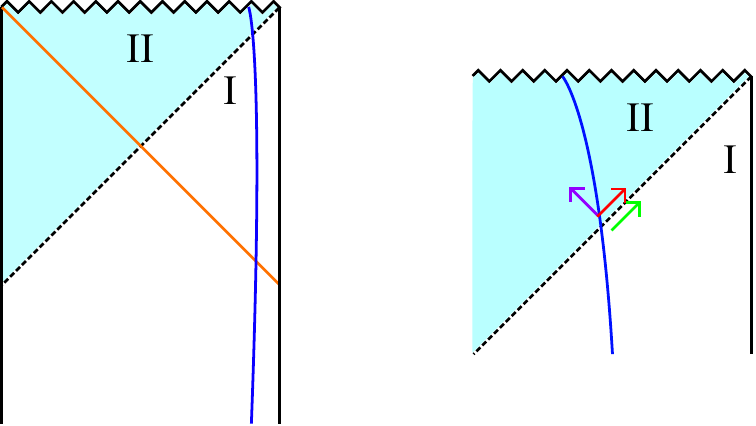}
\caption{A one-sided AdS black hole.  On the left we have the shell of matter that created the black hole in orange, the black hole interior in light blue, the horizon as a dashed line, and an infalling observer in dark blue.  On the right is a detail of the region where the observer crosses the horizon, with an interesting set of modes indicated.  The purple modes are easily evolved back to region I, and the green modes are already there.  The red modes would need to be evolved all the way back through the collapsing shell and reflected off of $r=0$ to get them out to region I.  Smoothness of the horizon requires entanglement between the red and green modes.}\label{adsinfall}
\end{center}
\end{figure}

Consider a big one-sided AdS black hole, made from some sort of infalling matter at early times.  The Penrose diagram for this system is shown in figure \ref{adsinfall}.  
Bulk effective field theory degrees of freedom in the region outside of the horizon, which I have denoted region I, can be fairly simply described in terms of microscopic CFT variables using the BDHM/HKLL map \cite{Banks:1998dd,Hamilton:2006az,Kabat:2011rz}.  This construction essentially proceeds by solving the bulk operator equations of motion in from the boundary in the CFT \cite{Heemskerk:2012mn}; I review a few more details in the following subsection.

To describe the interior, denoted as region II in the figure, is more challenging.  One way to begin is to observe that the interior lies to the future of everything outside, so roughly we can think of the horizon as a Cauchy surface and then evolve up into region II using the bulk equations of motion \cite{Freivogel:2004rd,Horowitz:2009wm,Heemskerk:2012mn}.  Left-moving modes just inside the horizon, shown in purple in figure \ref{adsinfall}, can indeed be simply understood as having ``just fallen in'' from region I.\footnote{This decomposition into left- and right-moving modes is made only in the vicinity of the horizon; because of mass, tranverse momentum, and/or interactions it will not be conserved globally.  For brevity I will sometimes ignore this in the following heuristic discussion; mixing can systematically be included without affecting the main points here.}  Right-moving modes behind the horizon however, which are shown in red in figure \ref{adsinfall}, are more subtle.  If we try to evolve them back, they are more and more blue-shifted and eventually collide with the infalling matter at high center of mass energy.  At this point the bulk equations of motion are insufficent to proceed further, and we are unable to reflect through $r=0$ and back out to find a simple CFT definition of these modes \cite{Almheiri:2013hfa}.

\begin{figure}
\begin{center}
\includegraphics[height=5cm]{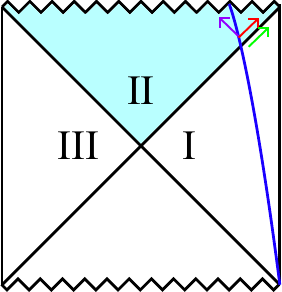}
\caption{The two-sided AdS-Schwarzschild wormhole.  The infalling observer again jumps in from the right, but the red right-moving modes inside can now be understood as having come from the left side.}\label{ads2side}
\end{center}
\end{figure}

This situation can be compared with the two-sided AdS-Schwarzschild wormhole, shown in figure \ref{ads2side}.  
The main difference for our purposes is that in figure \ref{ads2side} the red modes can be evolved back to the left boundary without encountering any high-energy collision.  This suggests that a CFT description of region II in the two-sided case should be easier than for the one-sided case; a construction along the lines of BDHM/HKLL should be possible \cite{Banks:1998dd,Hamilton:2006az,Kabat:2011rz}, and indeed some of the details have been worked out in \cite{Papadodimas:2012aq}.  In order to proceed similarly for the one-sided black hole the task then seems clear: where are the red modes in the CFT?

\subsection{The Basics of Reconstruction}
I'll first review a bit more about the BDHM/HKLL construction of local bulk operators in AdS/CFT \cite{Banks:1998dd,Hamilton:2006az,Kabat:2011rz}.  Any such construction will at best be perturbative in the gravitational coupling constant, which I will refer to as $1/N$.  For small numbers of operators any backreaction can be treated perturbatively, so by an appropriate gauge fixing \cite{Heemskerk:2012np} we can treat the bulk theory as a quantum field theory in curved spacetime (the gravitons will just be another matter field).  It will be an effective field theory with nontrivial irrelevant operators appearing that are suppressed by powers of $1/N$; their coefficients can in principle be determined by comparison with the CFT.  Which background we use depends on the classical properties of the state under consideration, such as its mass and charge.  For simplicity I will assume that all interactions are suppressed by powers of $1/N$, as for example would be the case in the asymptotically $AdS_4 \times \mathbb{S}^7$ superselection sector of M-theory that is dual to the ABJM theory \cite{Aharony:2008ug}.  

To leading order in $1/N$ all fields are then free, so in particular for a massive scalar field we have the Heisenberg picture expression
\be\label{bulkphi}
\phi(x)=\sum_n\left[f_n(x)a_n+f_n^*(x)a_n^\dagger\right],
\ee
where $f_n$ are Klein-Gordon normalizeable solutions of the bulk wave equation in the background of interest and $a_n$ and $a_n^\dagger$ are annihilation and creation operators obeying the usual algebra.  I will always consider the CFT on $\mathbb{R}\times \mathbb{S}^{d-1}$, so the index $n$ will be discrete.  Moreover I will always take the modes $f_n$ to have definite angular momentum and positive ADM energy, meaning that they will approach $r^{-\Delta}e^{-i\omega t}Y_{\ell m_1\ldots m_{d-2}}(\Omega)$, with $\omega\geq0$,  at large $r$ in coordinates where the metric approaches
\be
ds^2=-(r^2+1)dt^2+\frac{dr^2}{r^2+1}+r^2 d\Omega_{d-1}^2.
\ee
Here as usual 
\be
\Delta=\frac{d}{2}+\frac{1}{2}\sqrt{d^2+4m^2},
\ee
and I have set the AdS radius to one.  In order for \eqref{bulkphi} to make sense as an operator expression in the CFT we need to give a CFT expression for $a_n$.  The right choice \cite{Banks:1998dd} turns out to be to take 
\be\label{opmap}
a_n\propto \mO_{\omega \ell m_1\ldots m_{d-2}},
\ee
where on the right hand side we have the Fourier transform 
\be
\mO_{\omega \ell m_1\ldots m_{d-2}}=\int dt \int d\Omega e^{i\omega t}Y^*_{\ell m_1\ldots m_{d-2}}(\Omega)\mO(t,\Omega),
\ee
where $\mO$ is the CFT primary operator dual to $\phi$.  For compactness I will simply refer to these operators as $\mO_\omega$ below; it will always be implicit that $\omega\geq 0$.  \eqref{opmap} is uniquely determined by requiring that \eqref{bulkphi} obey its bulk equation of motion and be consistent with the ``extrapolate'' dictionary \cite{Banks:1998dd,Harlow:2011ke}
\be
\lim_{r\to\infty}\phi(t,r,\Omega)r^{\Delta}=\mO(t,\Omega).
\ee
One can check that $\mO_\omega$ and $\mO_\omega^\dagger$ have the right algebra to leading order in $1/N$, this follows for example from the large $N$ operator product expansion
\be
\mO(y)\mO(y')=\frac{1}{|y-y'|^{2\Delta}}+\mO^2(y')+O(1/N).
\ee
Here I use $y$ to denote a boundary point, as opposed to $x$, which is a bulk point.  

One can also write \eqref{opmap} in position space \cite{Hamilton:2006az,Kabat:2011rz} as
\be\label{position space}
\phi(x)=\int d^d y \sqrt{\gamma(y)}K(x;y)\mO(y),
\ee 
where $\gamma$ is the boundary metric and $K(x;y)$ is sometimes called the ``smearing function''.\footnote{For some backgrounds this expression does not quite exist as written \cite{Leichenauer:2013kaa}, but it can always be fixed up by smearing the bulk operator a little.}  This position space expression is convenient in the treatment of $1/N$ corrections \cite{Heemskerk:2012mn}. For example in the presence of a cubic interaction $\frac{g}{3N}\phi^3$, solving the bulk equation of motion to next to leading order in $1/N$ gives \cite{Heemskerk:2012mn}
\begin{align}\nonumber
\phi(x)=&\int d^d y \sqrt{\gamma(y)}K(x;y)\mO(y)\\\nonumber
&+\frac{g}{N}\int d^{d+1} x'\sqrt{-g(x')} d^d y \sqrt{\gamma(y)}d^d y' \sqrt{\gamma(y')}G(x;x')K(x';y)K(x';y')\mO(y)\mO(y')\\\label{mapcorr}
&+O(1/N^2).
\end{align}
Here $G$ is a particular type of bulk Green's function. The right hand side has an obvious diagrammatic interpretation that continues to higher orders.  We will not need the details though, the point for us is just that the right hand side involves (nonlocal) polynomials of higher and higher order in $\mO_\omega$ as we go to higher order in $1/N$.  The $\mO_\omega$'s are thus the ``building blocks'' one uses to perturbatively assemble bulk fields.

In the CFT the $\mO_\omega$'s are somewhat singular operators, for example they exactly obey 
\be
[H,\mO_\omega]=-\omega \mO_\omega,
\ee
so they have nonzero matrix elements only between energy eigenstates that differ by exactly $\omega$.  Papadodimas and Raju suggest redefining them by integrating over a small frequency range to make them more robust \cite{Papadodimas:2013wnh,Papadodimas:2013jku}; I will instead leave them as they are but insist on using them only in wave packets that are localized to within some time range $\Delta t$.

\subsection{The Two-sided Bulk}
Let's now consider interacting bulk fields propagating on the two-sided geometry of figure \ref{ads2side}.  There is a natural CPT transformation $\Theta$ which exchanges fields on the two sides; in Schwarzschild coordinates we have the action
\be
\Theta^\dagger \phi_{I}(t,r,\Omega)\Theta= \phi_{III}(-t,r,\Omega),
\ee
where $\phi$ is a real bulk scalar field.  The Hilbert space on the slice $t=0$ is a tensor product of states in region I and states in region III, each of which is conveniently spanned by eigenstates of the Hamiltonians $H_R$ and $H_L$ respectively.\footnote{This decomposition and the definition of $H_R$ and $H_L$ are straightforward for scalars, but for gauge fields and gravity there is some subtlety due to the constraints at the interface between regions I and region III \cite{Donnelly:2011hn,Donnelly:2012st,Casini:2013rba,Radicevic:2014kqa}.  I discuss this at some length in appendix \ref{gaugeapp}, but the upshot is that I do not expect these subtleties to affect equations \eqref{HOcom}, \eqref{HOcom2}, and \eqref{bulkmir} below, with $H_R$ and $H_L$ interpreted as the ADM Hamiltonians.}  It is convenient to define $\Theta$ as a map from the left Hilbert space to the right Hilbert space rather than from the full Hilbert space to itself, for example we can then use a basis $|i\ran_R$ of $H_R$ eigenstates for the states in region I and a basis
\be
|i^*\ran_L \equiv \Theta^\dagger |i\ran_R
\ee
of $H_L$ eigenstates for the states in region III.  An operator $A$ in region I with action
\be
A|i\ran_R=\sum_j A_{ji}|j\ran_R
\ee
has a $CPT$ conjugate which acts as
\be\label{CPTconj}
\Theta^\dagger A \Theta|i^*\ran_L=\sum_j A^*_{ji}|j^*\ran_L.
\ee
We will be interested in operators $\mO_\omega$ on the right which obey
\begin{align}\nonumber
[H_R,\mO_\omega]&=-\omega \mO_\omega\\
[H_L,\mO_\omega]&=0\label{HOcom}
\end{align}
Their CPT conjugates will then obey
\begin{align}\nonumber
[H_L,\Theta^\dagger \mO_\omega \Theta]&=-\omega \mO_\omega\\
[H_R,\Theta^\dagger \mO_\omega \Theta]&=0.\label{HOcom2}
\end{align}
Intuitively $\mO_\omega$'s create and annihilate excitations with $H_R-H_L=\omega$ in region I, their CPT conjugates create and annihilate excitations with $H_R-H_L=-\omega$ in region III, and both are needed to understand region II (or IV).  Up to mixing  the purple modes in figure \ref{ads2side} are created and annihilated by acting with $\mO$'s and time-evolving, while the red modes are created by acting with their CPT conjugates and evolving. 

Let's now consider a bulk state
\be\label{psib}
|\psi_{bulk}\ran\equiv \sum_{ij}C_{ji}|i^*\ran_L |j\ran_R.
\ee
If $C$ is invertible then this state has the interesting property that
\be
\Theta^\dagger A \Theta|\psi_{bulk}\ran=C A^\dagger C^{-1}|\psi_{bulk}\ran.  
\ee
In other words, the action of an operator in region III on the state can be written (in a way that depends on the state) as an operator acting on region I.  That this is possible is a consequence of the entanglement of the state \ref{psib}; it is the same basic idea that is at the heart of the Reeh-Schlieder theorem of relativistic quantum field theory \cite{Streater:1989vi}.  A similar equation holds for the action of $\Theta^\dagger A \Theta$ on a more general state obtained by acting on the state \eqref{psib} with another operator $A'$ on the right:
\be\label{bulkmir}
(\Theta^\dagger A \Theta) A'|\psi_{bulk}\ran=A'(C A^\dagger C^{-1})|\psi_{bulk}\ran.
\ee

One interesting bulk state to consider is the Hartle-Hawking-Israel or thermofield double (TFD) state \cite{Hartle:1976tp,Israel:1976ur} 
\be\label{TFD}
C_{ji}\propto \delta_{ij} e^{-\beta E_i/2},
\ee
which is the natural choice of ``ground state'' for the system.  In this case an operator $\mO_\omega$ obeying \eqref{HOcom} will have a CPT conjugate that acts on the TFD state as
\be
\Theta^\dagger \mO_\omega \Theta|\psi_{bulk}\ran=e^{-\beta\omega/2}\mO_\omega^\dagger|\psi_{bulk}\ran.
\ee
In AdS/CFT the TFD state is dual to itself, with the bulk energy eigenstates replaced by energy eigenstates of two copies of the CFT \cite{Maldacena:2001kr}.  We will be interested in more general states than the TFD, so for the most part we will stick to the general expression \eqref{psib}, assuming only that $C$ is invertible. The reduced density matrix on the right is then
\be\label{rhoR}
\rho_R=CC^\dagger.
\ee

\subsection{The Papadodimas-Raju proposal}
The proposal of Papadodimas and Raju is to use equation \eqref{bulkmir} to motivate a construction of the red modes in figure \ref{adsinfall}; if the extra operators from the left side that we need in the two-sided case to construct the interior can be rewritten as operators acting on the right side, why don't we just find CFT operators with this action and use them in the one-sided case as well?  There are several issues with trying to do this however.

\begin{figure}
\begin{center}
\includegraphics[height=5cm]{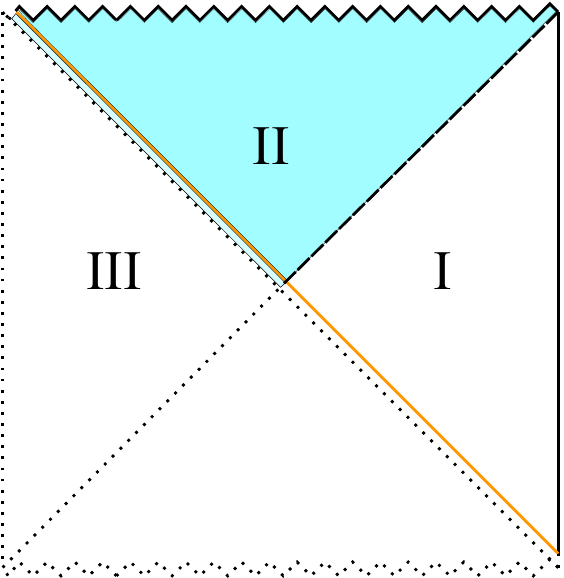}
\caption{The ``method of images''.  We define operators in region III, but use them only to compute things which are localized above the orange shell.  The expression of operators in region II in terms of the region III (and I) operators is found by using the two-sided \textit{bulk} equations of motion, \textit{assuming that the shell does not exist}.  Of course the shell does exist, its existence can be confirmed just in region I using the ordinary AdS/CFT dictionary, and the regions below it in this figure therefore do not.}\label{ghost}
\end{center}
\end{figure}

First of all any attempt to produce the $\Theta^\dagger A \Theta$ operators in the single CFT of a one-sided collapse seems like it will accomplish too much: in addition to constructing region II we will also construct region III.  This would be overkill; if we really have region III then we should also have a second copy of the CFT.  The situation here however is similar to the method of images in electrostatics; we use the $\Theta^\dagger A \Theta$ 's \textit{only} to compute things in region II.  This is illustrated in figure \ref{ghost}.

Secondly \eqref{bulkmir} depends on the matrix $C$, which came from the choice of bulk state \eqref{psib}.  So if $C$ appears in our definition of operators behind the horizon, we are essentially putting in the state that we want to get out.  But how do we choose it?  For now I will just assume that there is some prescription which in some appropriate sense agrees with the TFD state to leading order in $1/N$; I return to this question in section \ref{TFDsec}. 

There is also an immediate technical problem.  Say we want to define a ``mirror'' operator $\tO_\omega$ in the CFT whose bulk interpretation is the same as $\Theta^\dagger \mO_\omega \Theta$ when used in constructing operators in region II.  Moreover say we have some ``smooth horizon'' finite-energy state $|\psi\ran$ in the CFT on which $\tO_\omega$ acts as $C\mO^\dagger_\omega C^{-1}$.  The full set of CFT states can be generated from this state by acting with enough local CFT operators \cite{Streater:1989vi}, but if we demand that $\tO_\omega$ acts as in equation \eqref{bulkmir} for $A'$ any polynomial of local CFT operators, then we would discover that $\tO_\omega$ commutes with all local operators and is thus proportional to the identity.\footnote{This is a version of the ``commutator'' argument of \cite{Almheiri:2013hfa}, which is a standard criticism of ``$A=R_B$''.}  One of the two main new ideas of the PR papers \cite{Papadodimas:2013wnh,Papadodimas:2013jku} is to ameliorate this by requiring $\tO_\omega$ to act as \eqref{bulkmir} only when $A'$ is in some small set of operators $\A$.   

More precisely, they define the set of operators $\A$ as the set of all polynomials in the $\mO_\omega$'s, their hermitian adjoints, the $C$ conjugates of both, the Hamiltonian, and the charges for any bulk gauge fields, with the restrictions that both the degree of the polynomial and the energy of all operators present cannot be too large.\footnote{The Hamiltonian and any conserved charges can be understood as zero modes of single-trace operators, but they are sufficiently special that they sometimes need to be discussed separately.  From here on the set of $\mO_\omega$'s and $\tO_\omega$'s should always be understood as \textit{not} including the zero modes of any conserved currents.  The symbol $A_\alpha$ will denote a generic element of the algebra $\A$, which does include them.}  I will denote the maximal degree as $d_{max}$, and require that the total energy be much less than the energy of the black hole.  I will also demand that each frequency index $\omega$ is integrated against a wave packet which localizes it to within some particular time range $\Delta t$.  

We can estimate the total number of linearly independent elements of the set $\A$ as follows.  First of all we can get the most operators for a given total energy by taking them all to have $\omega\lesssim \frac{1}{r_s}$, where $r_s$ is the Schwarzschild radius.\footnote{For simplicity I will occasionally assume that $r_s$ is not parametrically larger in $N$ than the AdS radius.  The temperature will then also be of order the AdS scale.}  To avoid the modes they create being confined to within a Planckian distance of the horizon we must cut off their total angular momentum at $\ell_{max}\sim \frac{r_s}{\ell_p}$.  The total number of angular momentum modes at a given frequency then scales like $\ell_{max}^{d-1}\sim S$.  The number of linearly independent wave packets at a given angular momentum is of order $\frac{\Delta t}{r_s}$, so we can estimate the number of linearly-independent elements of $\A$ as
\be
|\A|\sim \left(\frac{\Delta t}{r_s} S\right)^{d_{max}}.
\ee 
In order to have a chance at nontrivial $\tO_\omega$'s we need
\be\label{algbound}
|\A|\ll e^S,
\ee
which we can obtain in various ways depending on what we assume about $\Delta t$.  PR take $\Delta t\sim r_s e^{\sqrt{S}}$, which would then imply that we need $d_{max}$ to be at most $\sqrt{S}$, but we could also take $\Delta t \sim r_s S$, or even $\Delta t\sim r_s$, in which case we can have $d_{max}\sim S/\log S$.
It is convenient to take the time range the wave packets are localized in to be centered at $t=0$, which we can do without loss of generality by moving the shell back in time as in figure \ref{ghost}.

Intuitively the set $\A$ is supposed to be the ``set of observables outside the black hole that are easy for an infalling observer to measure''.\footnote{An important point here is that ``easy'' is different from ``possible''; things which involve $O(S)$ $\mO_\omega$'s seem quite possible to measure.  I return to this below.}  These restrictions mean that $\A$ is not quite a Von Neumann algebra, since it is not closed under multiplication.  I will nonetheless sometimes refer to it for convenience as the ``small algebra''.  

For any state $|\psi\ran$ in the CFT one can then define a subspace
\be
\mathcal{H}_\psi\equiv \A |\psi\ran.  
\ee
Inspired by \eqref{bulkmir}, PR then suggest defining the action of the mirror operators on $\mathcal{H}_\psi$ as
\begin{align}\nonumber
\tO_{\omega}|\psi\ran&=C \mO_\omega^\dagger C^{-1}|\psi\ran\\
\tO_{\omega}^\dagger|\psi\ran&=C \mO_\omega C^{-1}|\psi\ran,\label{mireq}
\end{align}
together with\footnote{In fact PR actually instead require that $\tO$ commutes only with $\mO$'s, while for the Hamiltonian $H$ they instead demand $[H,\tO_\omega]=\omega \tO_\omega$ (and a similar equation for any conserved charge $Q$).    I explain in appendix \ref{gaugeapp} why I prefer the prescription given here.  The difference comes from whether we interpret the CFT Hamiltonian $H$ as representing the bulk operator $H_R$ or the bulk operator $H_R-H_L+E_0$ with $E_0$ some constant.  My choice is the former, whereas they would like the latter, but only the former seems consistent with the OPE structure of the CFT.}
\be
[\tO_\omega,A_\alpha]\mathcal{H}_\psi=[\tO_\omega^\dagger,A_\alpha]\mathcal{H}_\psi=0.\label{algeq}
\ee
Equations \eqref{mireq} say that the mirror operators act on the state $|\psi\ran$ as if they were $\Theta^\dagger \tO\Theta$ acting on the bulk state \eqref{psib}, and equations \eqref{algeq} say that their algebra acting on $\mathcal{H}_\psi$ is the same as it would be in the bulk.  In other words what the proposal is doing is ``simulating'' the two-sided bulk of the previous section within a single copy of the CFT.  These equations can be interpreted as a set of linear constraints which must be solved to find the mirror operators; they should be solveable provided they are consistent.\footnote{Their consistency essentially follows from \eqref{algbound} and the equilibrium condition \eqref{eqcond} below; for details see the PR papers \cite{Papadodimas:2013wnh,Papadodimas:2013jku}.} Note that \eqref{mireq} and \eqref{algeq} are uncorrected to all orders in $1/N$; perturbative corrections all go into the map to bulk fields, as in equation \eqref{mapcorr}, and the choice of $C$.  

This definition immediately begs the question however of which states $|\psi\ran$ should be used to define $\mathcal{H}_\psi$.  The second main new idea of the PR papers \cite{Papadodimas:2013wnh,Papadodimas:2013jku} is to give a rule for which states $|\psi\ran$ should be used.  The idea is that we should only expect \eqref{mireq} to be satisfied if the state $|\psi\ran$ is an ``equilibrium'' state.  There are various ways to define the equilibrium condition, the one I will mostly use is that in any equilibrium state $|\psi\ran$ we must have\footnote{This is different than equilibrium condition proposed by PR; they demand only that expectation values of elements of $\A$ are time-independent to exponential precision.  This is not sufficient however for the correlation functions of $\mO$'s and $\tO$'s to reproduce bulk correlators in the state \eqref{psib}.  For example a superposition of two black holes of very different mass would be an equilibrium state according to their criterion, since no elements of $\A$ mix between them.  My condition \eqref{eqcond} implies theirs when $C$ commutes with the Hamiltonian, but is also necessary and probably sufficient for the consistency of the proposal.}
\be\label{eqcond}
\lan\psi|A_\alpha |\psi\ran=\mathrm{tr} (C C^\dagger A_\alpha)+O\left(e^{-c S}\right)
\ee
for any $A_\alpha\in \A$, for some $O(1)$ constant $c$.  This condition has two motivations; first of all we certainly shouldn't expect a CFT state $|\psi\ran$ to look like the bulk state \eqref{psib} (evolved up to region II) unless the expectation values of operators in region I constructed by the BDHM/HKLL map are consistent with this.  This map is supposed to accurately reconstruct the bulk to all orders in $1/N$, so any differences should be non-perturbatively small.\footnote{Remember we are considering big black holes so $S$ is proportional to some positive power of $N$.}  Secondly, since we assumed that to leading order in $1/N$ we have $C C^\dagger\approx \frac{1}{Z}e^{-\beta H_{CFT}}$, where we can now define this approximation more carefully as meaning that the expectation values of elements of $\A$ in the two ensembles agree to leading order in $1/N$, we can think of states obeying \eqref{eqcond} as being states where the black hole has ``settled down'' enough that the state looks thermal with respect to the small algebra $\A$.  In particular any objects would have to have been thrown in more than a time of order $r_s S$ in the past in order for the excitations they created to die down to exponentially small size.\footnote{From the point of view of this observation it seems rather natural to take $\Delta t\sim r_s S$, since this gives the infalling observer the ability to do experiments involving equilibration to the level of precision involved in \eqref{eqcond}.  Having $\Delta t$ be shorter, for example of order $r_s \log S$, seems too restrictive given our intuition that $\A$ should represent what is ``not too hard'' to do.}  By the argument of Hayden and Preskill \cite{Hayden:2007cs} it is then much too late for them to affect the experience of an observer who jumps in near $t=0$.  Thus the equilibrium states ``have a right'' to a smooth horizon.  By using equations \eqref{mireq}, \eqref{algeq}, and \eqref{eqcond}, it is clear that any expectation value of bulk fields constructed from the $\mO$'s in the small algebra $\A$ and their mirror $\tO$'s will agree with low energy effective field theory in the state \eqref{psib} to all orders in $1/N$.

Equilibrium states obey an important ``KMS'' condition
\be\label{KMS}
\lan\psi|A_\alpha A_\beta|\psi\ran=\lan\psi|A_\beta C C^\dagger A_\alpha (C C^\dagger)^{-1}|\psi\ran+O\left(e^{-cS}\right),
\ee
for any two elements $A_\alpha$, $A_\beta$ in $\A$.  This condition is necessary for the consistency of \eqref{mireq}, since it ensures that the action of $\tO_\omega^\dagger$ on the right is consistent with its natural action on the left induced from the action of $\tO_\omega$ on the right.\footnote{I thank Herman Verlinde and Xi Dong for discussions of this point.}

The $\tO$ operators have the uncomfortable property that they are ``state-dependent''; ordinarily in quantum mechanics one first defines an observable to be some hermitian operator and then sticks to this hermitian operator regardless of the state of the system.  For now we will just accept this, but I will give a detailed discussion of to what extent this is a modification of quantum mechanics (it is) in section \ref{sdsec}.

\subsection{Choosing the bulk state}\label{TFDsec}
I now return to the choice of the ``target'' bulk state \eqref{psib}.  We should really think of the equilibrium condition \eqref{eqcond} as a ``compatability'' condition between a set of CFT equilibrium states $\mathcal{E}$ and a two-sided bulk state labelled by $C$; in order to realize the PR proposal we must look for compatible pairs $(\mathcal{E},C)$.  The most obvious $C$ to consider is the TFD state, and a set of CFT states which is compatible with it is
\be\label{thermalpure}
|\psi\ran=\frac{1}{\sqrt{Z}}\sum_j e^{-\beta E_j/2+i\phi_j}|j\ran,
\ee
where $|j\ran$ are energy eigenstates and $\phi_j$ are randomly chosen phases.  The compatibility, meaning that expectation values of elements of $\A$ obey \eqref{eqcond} with $CC^\dagger=\frac{1}{Z}e^{-\beta H}$, follows from the eigenstate thermalization hypothesis \cite{Srednicki:1995pt}
\be
A_{ij}=\delta_{ij}A(E_i)+e^{-S\left(\frac{E_i+E_j}{2}\right)/2}R_{ij},
\ee  
where $A$ and $|R|$ are smooth $O(1)$ functions of $E$ but the phase of $R$ varies erratically.  The states \eqref{thermalpure} are thus a very promising starting point for implementing the PR proposal.

Interestingly however at the level of $1/N$ corrections other natural sets of ``black hole-like'' CFT states are \textit{not} compatible with the TFD state; this was briefly pointed out by PR \cite{Papadodimas:2013jku}.  I now discuss the reason for this at some more length. 
Consider a black hole formed in a pure state from a collapse that lies in some thin energy shell.  If the state is sufficiently generic, expectation values of simple operators should be exponentially close (in the entropy) to their \textit{microcanonical} expectation values.  This follows for example from a theorem of Lloyd \cite{lloydthesis}, which states that for any operator $A$ on a Hilbert space of dimension $d$, we have
\be\label{lloydform}
\int dU \left(\lan\psi(U)|A|\psi(U)\ran-\lan A\ran_{MM}\right)^2=\frac{1}{d+1}\left(\lan A^2\ran_{MM}-\lan A\ran_{MM}^2\right),
\ee
where $MM$ denotes the expectation value in the maximally mixed density matrix $I/d$.  Here $|\psi(U)\ran$ denotes the state created by acting on some reference state with a unitary matrix $U$, which is then integrated over the Haar measure.  In other words the expectation value of any operator in a particular pure state is generically exponentially close (in the entropy $\log d$) to its maximally mixed expectation value.\footnote{Here I assume that $A$ is sufficiently smooth that $A^2$ does not have an expectation value which is exponentially enhanced; this should be the case for any operators we consider here.}  We'd like to take this Hilbert space to be the set of CFT states in some narrow energy band, ie the microcanonical ensemble, but we have the technical issue that not all operators in the small algebra $\A$ send this subspace into itself.  For any operator $A_\alpha\in \A$ however we can always construct an operator that \textit{does} send the microcanonical subspace into itself by just sandwiching $A_\alpha$ between two projection operators that project onto the subspace.  Lloyd's result \eqref{lloydform} applies to this projected operator, but actually we can ignore the projections in three of the four terms.  Indeed we have
\be\label{lloyd2}
\int dU \left(\lan\psi(U)|A_\alpha|\psi(U)\ran-\lan A_\alpha\ran_{MC}\right)^2=\frac{1}{d+1}\left(\lan A_\alpha\Pi A_\alpha\ran_{MC}-\lan A_\alpha\ran_{MC}^2\right),
\ee
where the average is now over pure states in the energy shell, $\Pi$ is the projection operator onto states in the shell, and $MC$ means the expectation value in the microcanonical density matrix that is proportional to the identity on this energy shell and is zero otherwise.  Finally if we take $A_\alpha$ to be hermitian then by inserting a complete set of energy eigenstates we see that 
\be
\lan A_\alpha\Pi A_\alpha\ran_{MC}\leq \lan A_\alpha^2\ran_{MC},
\ee  
which together with \eqref{lloyd2} immediately shows that the expectation value of a reasonably smooth operator in a typical pure state drawn from the microcanonical ensemble will be exponentially close to its microcanonical expectation value.\footnote{One might worry that these Haar averaged states are ``too typical'' in the sense that they usually must be built up over exponentially long times.  In fact Lloyd's theorem holds for averages over much simpler sets of states, such as those generated by unitary 2-designs \cite{Hayden:2007cs}.}  

For comparison we can study how accurately the canonical ensemble reproduces the expectation values of elements of $\A$ in a collapse state $|\psi\ran$ of narrow energy width.  Expectation values in the canonical ensemble tend to differ from those in the microcanonical ensemble by \textit{powers} in the inverse entropy, so in fact the TFD state will already get the expectation values wrong at low orders in perturbation theory.\footnote{It may be somewhat unfamiliar to see ensemble inequivalence competing with perturbation theory in interactions; the reason is that for a big black hole we have taken the entropy to be of order $N$ to some power while the interactions are suppressed by powers of $1/N$.  This is different than the usual situation in statistical mechanics where interactions are suppressed only by factors like $1/137$ while entropies are of order $10^{23}$.}  
Indeed for any operator $A_\alpha$ with reasonably smooth diagonal matrix elements in energy, we can estimate its canonical expectation value as
\be
\lan A_\alpha\ran=\frac{\int dE e^{S(E)-\beta E}A_\alpha(E)}{\int dE e^{S(E)-\beta E}}.
\ee 
The saddle point approximation to these integrals gives back the microcanonical expectation value, but the perturbative corrections to the saddle point will only be suppressed by powers of the entropy, which is not good enough to satisfy the equilibrium condition \eqref{eqcond}.

Thus we see that for black holes formed by a collapse that is well-localized in energy, it seems we should look for a ``target'' bulk state \eqref{psib} where the reduced density matrix $CC^\dagger$ is close to the \textit{microcanonical} density matrix, which is constant in some energy range and then very small outside of it.  This however is actually rather problematic from the point of view of the PR construction.  The obvious choice would be to take $C$ to be the ``microcanonical double state'', where $C$ is diagonal with real and positive elements.  But in this case the action of the mirror operators is rather badly defined; consider the state
\be
\tO_\omega|\psi\ran=C\mO_\omega^\dagger C^{-1}|\psi \ran,
\ee
where $\omega$ is parametrically larger than the width of the energy band from which we pull $|\psi\ran$.  The $C^{-1}$ keeps the state $|\psi\ran$ within the band, but the $\mO_\omega^\dagger$ takes it out.  When we then act with $C$ again we will then get a huge suppression, with an amount that depends on how exactly we define the microcanonical ensemble $CC^\dagger$ outside of the energy range we are interested in.  Thus if we compute a correlation function like $\lan \tO \mO\ran$ it will be exceedingly small.  This suggests that to the extent the state has a geometric interpretation at all,  it does not involve two sides which are separated only by a single bifurcate horizon.  It then is far from clear that the $\mO$'s and $\tO$'s provide sufficient initial data to reconstruct region II a la figure \ref{ghost}.\footnote{This argument does not apply to the TFD because its energy width is of order the temperature times $\sqrt{S}$, which is larger than $\omega$ for any operators of interest for the infalling observer.}  This inability to deal with narrow states is problematic for the generality of the construction, since after all one would hope that for example exact energy eigenstates should have smooth interiors, and in appendix \ref{hamsec} I give some brief speculation on what might be done about it.\footnote{This difficulty with states of narrow energy width is one of the main reasons that Raju and Papadodimas attempted to have $[H,\tO]\neq 0$.  If this were possible it would ameliorate the problem somewhat, but I argue in appendix \ref{hamsec} that it does not seem to be consistent within the CFT to do this.} For now to avoid this issue, which is something of a distraction from the main point of this paper, I will just restrict the discussion to bulk states where the energy fluctuations in $CC^\dagger$ are of order those in the TFD state.

\section{Do States have Unique Interpretations?}\label{nonusec}
Let's now try to understand better the global structure of the CFT Hilbert space in the PR proposal.  There is a set $\mathcal{E}\subset\mathcal{H}_{CFT}$ of equilibrium states satisfying \eqref{eqcond}, relative to each of which one defines mirror operators $\tO$ with respect to which it resembles the bulk state \eqref{psib} for any infalling observer who jumps in in the vicinity of $t=0$.  For observers who jump in much later or much earlier we use a different choice of the small algebra $\A$, so the set $\mathcal{E}$ is different.  The set $\mathcal{E}$ is \textit{not} a linear subspace of the Hilbert space; in fact its span (including different energies) is just $\mathcal{H}_{CFT}$.  On top of each equilibrium state $|\psi\ran$ we then build a linear subspace $\mathcal{H}_\psi$ by acting with either elements of $\A$ or their mirror operators.  The other states in this subspace are to be interpreted as ``excited'', in some particular way.  This leads to what seems to be an important consistency requirement for the proposal: the linear subspaces constructed in this way must not intersect. 

Say that there was a state $|\chi\ran$ in the Hilbert space which could be realized either by acting on some equilibrium state $|\psi\ran$ with an operator $A_\alpha\in \A$ \textit{or} by acting on some other equilibrium state $|\psi'\ran$ with a different operator $A_\beta\in \A$.  In this case the physical interpretation of the state $|\chi\ran$ would be ambiguous; would an infalling observer see it as acting on the bulk state \eqref{psib} with $A_\alpha$ or with $A_\beta$?  I will now argue that this situation can indeed be generically realized and thus that in the PR proposal quantum states in $\mathcal{H}_{CFT}$ cannot have fixed physical interpretations.

To demonstrate such a situation it is clearly sufficient to find a nontrivial element of the algebra $\A$ which sends equilibrium states to other equilibrium states.  It is not immediately clear that this can be done, after all the equilibrium condition \eqref{eqcond} is rather restrictive.  Acting with any small number of $\mO_\omega$'s and $\mO_\omega^\dagger$'s can always be detected by the expectation value of some other small number of $\mO_\omega$'s and $\mO_\omega^\dagger$'s, since we can always just arrange to have a non-vanishing correlation function.  What we would like is a unitary transformation $\wt{U}$ that commutes with everything in $\A$ to exponential accuracy: we then would have
\be
\lan\psi|\wt{U}^\dagger A_\alpha \wt{U}|\psi\ran=\lan\psi| A_\alpha|\psi\ran.
\ee
An obvious guess for how to find such a $\wt{U}$ is to build it out of $\tilde{O}$ operators, since from \eqref{algeq} these commute with everything in $\A$.  For example we can consider the operator 
\be\label{tUeq}
\wt{U}\equiv e^{i \alpha \tO_\omega^\dagger \tO_\omega},
\ee
At leading order in $1/N$ this is the exponential of the number operator for some mode behind the horizon; it rotates the phases of the number eigenstates for the mode.   At higher order in $1/N$ it does not exactly have this interpretation, but it is well defined and from \eqref{algeq} it continues to commute with everything in $\A$ acting on the state $|\psi\ran$.\footnote{If we could arrange $[H,\tO]\neq 0$ as advocated by PR, then here we would need to arrange for $\wt{U}$ to commute with $H$ within expectation values.  This is rather restrictive, but seems to be possible by systematically ``improving'' \eqref{tUeq}.  I argue in appendix \ref{hamsec} however that we must have $[H,\tO]=0$.}  This operator thus sends the equilibrium state $|\psi\ran$ to another equilibrium state according to \eqref{eqcond}, but according to bulk effective field theory the horizon is no longer smooth.  More precisely the state $\wt{U}|\psi\ran$ is no longer annihilated by the ``infalling'' annihilation operator proportional to $\tO_\omega-C \mO_\omega^\dagger C^{-1}$, and in fact the ``infalling'' number operator has an $O(1)$ expectation value.  

The operator $\wt{U}$ is not actually an element of $\A$ since it involves the mirror operators, but we can use \eqref{mireq} to define a new operator that has the same action on $|\psi\ran$:\footnote{Technically for this equation to be valid we must perform the truncation of the exponential discussed in the following paragraph.}
\be
V\equiv e^{i \alpha C \mO_\omega^\dagger \mO_\omega C^{-1}}= C U^\dagger C^{-1},
\ee
where $U\equiv e^{-i\alpha \mO_\omega^\dagger \mO_\omega}$ is the unitary operator whose mirror is $\wt{U}$.  $V$ is not unitary, but its action on $|\psi\ran$ preserves the norm since it is equivalent to the action of $\wt{U}$.  It may appear surprising that acting with $V$ on the state preserves the expectation values of all elements of $\A$, but this amusingly follows from the KMS condition \eqref{KMS}.  Indeed
\begin{align}\nonumber
\lan\psi|V^\dagger A_\alpha V|\psi\ran&=\lan\psi|(C^{\dagger})^{-1}U C^\dagger A_\alpha C U^\dagger C^{-1}|\psi\ran\\\nonumber
&=\lan\psi|A_\alpha C U^\dagger C^{-1}C C^\dagger (C^\dagger)^{-1}U C^\dagger (C^\dagger)^{-1}C^{-1}|\psi\ran+O(e^{-cS})\\
&=\lan\psi|A_\alpha|\psi\ran+O(e^{-cS}).
\end{align}
This argument applies for any $\wt{U}$ that we build out of $\tO$'s.  

To complete the argument we now would like to argue that $V\in \A$, but this isn't actually true, for two reasons. First of all it is not a polynomial in $\mO_\omega$, $\mO_\omega^\dagger$, and their $C$ conjugates of degree at most $d_{max}$.  Secondly we are supposed to integrate any $\omega$ index against a wave packet that localizes it into a time range $\Delta t$.  The wave packets are easily included, and to deal with the first problem the convergent series expansion for the exponential in the definition of $V$ can simply be truncated at order $d_{max}$.  This breaks the unitarity of $U$, but only by an amount which is of order $e^{-d_{max} \log d_{max}+\# d_{max}}$, where $\#$ is some $O(1)$ number.\footnote{Here I have assumed that the we can think of the operator $C \mO_\omega^\dagger \mO_\omega C^{-1}$ as being bounded at order one.  Since to leading order in $1/N$ it is just a number operator, this will clearly be true for a fermionic field.  For a bosonic field, we need to use the property that eigenstates of the number operator with large eigenvalue are exponentially suppressed in an equilibrium state.}  If we take $\Delta t$ to be at most some power of $S$, then we found in the discussion around \eqref{algbound} that we can consistently take $d_{max} \sim S/\log S$; the error is then of order $e^{-c S}$, which doesn't violate the equilibrium condition.

If we take $\Delta t$ to be of order $e^{\sqrt{S}}$, then we can only make the error as small as $e^{-\sqrt{S}}$ (the actual choice of power here is unimportant, I take $1/2$ for simplicity of exposition).  This is non-perturbatively small, but still parametrically larger than $e^{-cS}$. Is a deviation from the equilibrium condition of order $e^{-\sqrt{S}}$ ``large enough'' to no longer expect a smooth horizon?   By the rules I've described so far it is, but there is some question as to whether or not it is really reasonable to insist on the equilibrium condition \eqref{eqcond} being so strong. Saying that the deviation is of order $e^{-c S}$ is a stronger statement that saying that it is non-perturbatively small; for example in string theory in the early 1990's it was a major accomplishment to realize that nonperturbative effects should scale like $e^{-1/g}$ instead of $e^{-1/g^2}$ \cite{Shenker:1990uf}.  As I discussed in the previous section however, even in perturbation theory it is unclear whether or not a well defined procedure exists for determining the target bulk state \eqref{psib}.  Beyond perturbation theory it is even less clear.  Consider for example the non-perturbative process where a black hole of mass of order the Planck mass is spontaneously fluctuated out of the horizon and into the atmosphere.  This decreases the entropy of the black hole by $S^{\frac{1}{d-1}}$, so we expect the probability of it happening is $e^{-S^{\frac{1}{d-1}}}$.  For $AdS_4$, this is $e^{-\sqrt{S}}$.  So apparently there are interesting non-perturbative effects of this size, which would be difficult to systematically include in determining the state \eqref{psib}, and since the matrix $C$ appears explicitly in the equilibrium condition it seems excessive to demand it to require such small deviations.   Of course even if we do require this the issue only arises if we take $\Delta t$ to be exponentially large, and there is no clear reason why we should do this.\footnote{Another possible loophole to the argument of this section is that we could simply \textit{declare} that $d_{max}$ is parametrically smaller than it needs to be for mirror operators to be consistently defined.  Since we are just making it up the rules anyway, there is no deep principle preventing this.  As long as we take it to scale like some power of $S$ however, the error from truncating the exponential in defining $V$ will be exponentially small in that power of $S$.  If the power is less than one then the caveats of this paragraph can again be applied to resist viewing this as a real resolution of the problem.  In any event making the algebra smaller than necessary is unsatisfying, since it is increasing in size the set of experiments which are in principle doable but not described by the PR proposal.}

We thus appear to have found a problem for the PR proposal; what is the bulk interpretation of the state $V|\psi\ran$?  Is the horizon excited or is it not?  In fact this issue is somewhat related to the difficulty in identifying the right target state \eqref{psib}; who is to say that we shouldn't include some extra phases in $C$?  Or even a generic unitary $\wt{U}$?  In fact if we were sufficiently perverse, we could make what seems to be an equally consistent version of the PR proposal where the mirror operators are defined in such a way that equilibrium states \textit{always} have firewalls.  The operators $\wt{U}$ are something like a ``zero mode'' that pushes us in the direction of such a definition.

\section{More General Black Holes}\label{genbhsec}
Having introduced and analyzed the PR proposal in the case where it is strongest, the big AdS black hole, I now discuss two more general cases which introduce new issues.  Another interesting generalization which I will not discuss is to big AdS black holes in states that are slightly mixed \cite{Verlinde:2013qya}. 

\subsection{Two-sided black holes}
I first consider two-sided AdS black holes.  The TFD state is obviously an interesting choice of state, where the interior seems to be describable in the BDHM/HKLL formalism without recourse to mirror operators.  We can also consider more general entangled states of the two CFT's, which should be dual to more complicated wormholes \cite{Maldacena:2013xja,Shenker:2013pqa,Shenker:2013yza}.  The new interesting question here is how the small algebra $\A$ should be defined.  Let's assume that the infalling observer will jump in from the right side; should $\A$ be given by its usual definition in the right CFT, or should it include ``simple'' operators from \textit{both} CFT's?  

Let's first imagine that we have $\A=\A_R$.  In the TFD state the mirror operators will then be the left algebra $\A_L$.  Any unitary operator acting on the left CFT preserves the equilibrium condition that operators in the right CFT have thermal expectation values, so in particular we could send in a freight train from the left boundary in figure \ref{ads2side} and it would not be detected by the small algebra $\A$.  In this setup it is thus even easier to get into the situation of the previous section; how do we know whether or not we should interpret the state with the freight train as a \textit{new} TFD state with a smooth horizon?  In fact the BDHM/HKLL construction here would say that we should not interpret it this way; the mapping between the left and the right CFT's and the bulk is fixed by the Euclidean construction of the TFD state, which connects the two sides in a single copy of the CFT; it does not leave any freedom to redefine the dictionary between the two sides.  So it seems taking the small algebra to just be $\A_R$ produces an inconsistency between the PR rules and the BDHM/HKLL dictionary.

We are thus led to consider the alternate choice of algebra where $\A$ is generated by the union of $\A_L$ and $\A_R$.  Here there is a new subtlety however; how should the equilibrium condition be defined?  One choice would be to require that all expectation values of elements of $\A$ resemble the TFD state.  In this case we would of course decide that the TFD state itself is an equilibrium state, but if we tried to construct mirror operators we would fail.  There would now be elements of $\A$ which annihilate the state, so the conditions \eqref{algeq} would not be solvable.  This is perhaps the correct answer, since in this case we do not need state-dependent mirror operators.  This choice however leads to a problem in that it seems incorrect when we consider more generic two-sided wormholes.  Let's consider a generic entangled pure state of two CFT's with fixed $H_R+H_L$.  According to the definition this would \textit{not} be an equilibrium state, since there would now not be any simple entanglement between the left and right algebras.  The PR construction would then not be able to tell us whether or not these states have smooth horizons, whereas if we don't believe in firewalls we might expect that they should.

We could instead use the two-sided algebra but define the equilibrium condition to be that the expectation values of the two-sided algebra are consistent with the product state
\be\label{2eq}
\rho_{Th}=\frac{e^{-\beta H_L}\otimes e^{-\beta H_R}}{Z^2}.
\ee
The typical two-sided pure state now will be an equilibrium state; the PR construction will produce two sets of mirror operators, one for each side, and it will construct a smooth horizon on each side.\footnote{In fact this will be the same construction as if we had just used the one-sided algebra for whichever side we jump in from.  This is reasonable, since generic wormholes are expected (if they are not singular!) to be ``long'' \cite{Shenker:2013pqa,Shenker:2013yza}; infallers from different sides won't be able to meet in the middle.}  The TFD state now will be far out of equilibrium, so the PR proposal will be silent on what its properties should be.  This is a good thing however, since as we just discussed we don't expect to need mirror operators to reconstruct the interior in the TFD state.  This last choice is thus probably the most appealing, even though for generic states it still will have two copies of the ambiguity of the previous section.

\subsection{Evaporating black holes}\label{evapsec}
I now turn to the evaporating black hole.  It is sometimes convenient to arrange for a big black hole in AdS to evaporate by coupling the CFT to an auxiliary system \cite{Rocha:2008fe,Almheiri:2013hfa}, but this can lead to puzzling issues which I would prefer to avoid so I will focus on an evaporating Minkowski black hole where we mostly expect local semiclassical physics to approximately hold everywhere.  The cost of course is that we cannot use the CFT language, so the discussion will be less precise.  

To be concrete I will model the state of an evaporating black hole as a qubit system, which factorizes into three parts \cite{Harlow:2013tf}
\be
\mathcal{H}=\HH\otimes\HB\otimes\HR.
\ee
Here $H$ is the remaining black hole (the ``stretched horizon''), which I take to consist of $m$ qubits, $B$ is the thermal atmosphere (``the zone''), which I take to have $k$ qubits, and $R$ is the Hawking radiation, which I take to have $n$ qubits.  We will mostly be interested in the situation where the black hole is ``old'', ie when $n>m+k$.  It is natural to take the small algebra $\A$ to be generated by polynomials in the Pauli operators acting on $\HB\otimes \HR$, since these are the degrees of freedom which are accessible to the infalling observer.  We will restrict to polynomials of degree at most $p$.  There is no natural dynamics in this model, so there is no analogue of the frequency wave packets we needed in the previous discussion.  Effectively we are just taking $\Delta t \sim r_s$.  We will consider a state $|\psi\ran$ to be an equilibrium state if 
\be
\lan\psi|A_\alpha|\psi\ran=2^{-n-m-k}\mathrm{tr} A_\alpha+O(2^{-c(n+m+k)}).
\ee
To implement a version of the PR proposal we need to pick a ``target'' state; we will imagine that the horizon is smooth in the infalling frame if we have
\be
|\psi\ran_{AB}=2^{-k/2}\sum_a|a\ran_A|a\ran_B,
\ee
where $a$ runs over $0$ and $1$ for each qubit.  $\HA$ is the ``image'' Hilbert space analogous to the second exterior in the PR proposal; we are essentially saying that each mode and its Hawking partner must be in the state $\frac{1}{\sqrt{2}}\left(|00\ran+|11\ran \right)$.  This state has the property that acting with the Pauli operator $Z_1$ on the first qubit has the same effect as acting with the Pauli operator $Z_2$ on the second qubit, and similarly for the Pauli $X_i$ operators and (up to a sign) the $Y_i$ operators.  This property is the analogue of equation \eqref{bulkmir} above.  We can then define mirror operators, for example by demanding that
\be\label{spinmir}
\wt{X}_i A_\alpha |\psi\ran=A_\alpha X_i |\psi\ran,
\ee
where $i$ runs over the qubits in $B$.  

It is interesting to see how large $p$ can be before we are no longer able to solve \eqref{spinmir} \cite{Papadodimas:2013jku}.  This happens when the set of states $A_\alpha|\psi\ran$ generated by acting on $|\psi\ran$ with linearly independent elements of the algebra stop being linearly independent.  The latest this can happen is when the number $|\A|$ of linearly independent elements of the algebra equals the dimensionality $2^{n+k+m}$ of the Hilbert space.  For the qubit system the linearly independent elements of the algebra are just products of Pauli matrices on the various sites, so if we include all products of degree at most $p$ then
\be
|\A|=\sum_{j=0}^p \begin{pmatrix}
n+k\\
j
\end{pmatrix}3^j.
\ee
If we take $p=n+k$ then this sum can be evaluated to give $2^{2(n+k)}$, as expected since in this case $\A$ would just be the set of all operators acting on $n+k$ qubits.  Mirror operators that commute with all operators on $B$ and $R$ can thus be defined only if $n+k<m$, or in other words if the black hole is ``young''.\footnote{This is a manifestation of Page's theorem, which says that when $m>n+k$ we can construct a purification of $B$ which lies entirely in $H$.  The mirror operators then can be defined to act only on $H$, so they manifestly commute with operators on $B$ and $R$.}  For old black holes however we clearly need to take $p<n+k$.  We can estimate how much less by defining $p\equiv (n+k)\alpha$, with $0<\alpha<1$, and approximating the sum as an integral: 
\be
|\A|\sim\int_0^\alpha d\alpha' e^{(-\alpha' \log \alpha'-(1-\alpha')\log(1-\alpha')+\alpha' \log 3)(n+k)}.
\ee
The integrand has a saddle point at $\alpha'=3/4$, where it is of order $2^{2(n+k)}$, so apparently when the black hole is old we need to take $\alpha<\frac{3}{4}$.  In that case it will be dominated by its upper endpoint, so we can determine the maximally allowed value by solving
\be
-\alpha \log \alpha-(1-\alpha)\log (1-\alpha)+\alpha \log 3=\log 2\left(1+\frac{m}{n+k}\right),
\ee
which is solved by some $O(1)$ value of $\alpha$ that is about $.2$ in the limit that $\frac{m}{n+k}\to 0$. 

We thus see that the PR proposal is able to arrange for the mirror operators to commute with any exterior operator that acts on at most $20\%$ of the atmosphere and Hawking radiation.  This is significantly more than was found in earlier attempts to make the idea of $A=R_B$ work, where it was typically found that the constructions of operators behind the horizon had $O(1)$ commutators even with single qubit operators on the radiation \cite{Almheiri:2013hfa,harlow}.  The reason the PR proposal is able to do so much better is that the $\wt{X}_i$ operators are not actually Pauli operators in the sense of having a spectrum which is half $1$ and half $-1$.  In other words there is no qubit factor in the Hilbert space on which they have the standard action; they are \textit{not} associated with some particular purification $R_B$ of $B$.  

Should we then be satisfied?  In this version of the PR proposal we can still raise the objections of section \ref{nonusec}, but I would instead like to draw attention to a different issue.  Namely, is it really reasonable to not allow a construction of the interior in situations where the infalling observer \textit{does} interact with more than $20\%$ of the Hawking radiation?  There seems to be no major technical obstruction to an infalling observer doing so, and the observer has plenty of time to do it before the black hole evaporates.\footnote{Such experiments are much easier than any experiment requiring decoding of the Hawking radiation, where there may indeed be a good case that the such experiments cannot be done by an observer who can also probe the interior \cite{Harlow:2013tf,aaronson}.}  Moreover even if the infalling observer does nothing, an $O(1)$ fraction of the radiation could interact with a dust cloud on its way out.\footnote{This objection was also raised by Raphael Bousso in a talk at the ``Bulk Microscopy from Holography'' workshop at the Princeton Center for Theoretical Science in November 2013.  See also \cite{Bousso:2013ifa}.}  Do we really expect a firewall in such situations?  I will postpone further discussion of this to section \ref{obsec} below, but I believe that any compelling theory of black hole physics will need to be able to describe such experiments, and any others that we can reasonably imagine doing.  It is not allowed to ``plead the fifth''.

\section{Some Comments on State-Dependence}\label{sdsec}
I now return to the question of state-dependence.  The goal of this section is to contrast the state-dependence of the PR proposal from more conventional phenomena which have something of the same flavor.  My basic strategy is to understand to what extent  ``state-dependent measurements'' can be realized as unitary evolution of the system to coupled to some apparatus.  I will argue that all ``standard examples'' of state-dependent measurement can be implemented in this way, but that the PR proposal cannot.  Before giving the general discussion it is convenient to first introduce an example.  
\subsection{State-Dependence and Spontaneous Symmetry Breaking}
Consider the $3+1$ dimensional $O(N)$ symmetric scalar field theory with Lagrangian
\be\label{ONlag}
\mathcal{L}=-\frac{1}{2}\partial_\mu \phi_i \partial^\mu \phi_i+\frac{m^2}{2}\phi_i \phi_i-\frac{g}{4}\left(\phi_i\phi_i\right)^2.
\ee
Here $i$ is an index which runs from $1$ to $N$, and the summation convention is in force.  In infinite volume (and with $m^2>0$) this theory has a continuous set of degenerate vacua $|\hat{n}\ran$, where $\hat{n}$ is a unit vector in $\mathbb{R}^N$.  These vacua can be distinguished by the expectation value of the field $\phi_i$, which at leading order in $g$ is
\be
\lan \phi_i(x)\ran_n\equiv \lan \hat{n}|\phi_i(x)|\hat{n}\ran=\frac{|m|}{\sqrt{g}}\hat{n}_i.  
\ee
The low-energy spectrum of this theory around one of the vacua has $N-1$ massless Goldstone bosons and one massive boson of mass squared $2m^2$.  

The point of interest for us here is that which fields create the Goldstone bosons seems to depend on the choice of state $|\hat{n}\ran$.  For example the Goldstone bosons are created by the field
\be\label{phistatedep}
\phi^\perp_{i}(x,\hat{n})\equiv \phi_i(x)-\left(\phi\cdot \hat{n}\right)\hat{n}_i.
\ee
Isn't this a state-dependent operator?  It appears to be, but before rushing to conclusions it is important to think more carefully about what we actually mean by ``measuring the two-point function of the Goldstone boson field''.  

One option is to try to ``remove'' the state-dependence by defining a single operator whose correlation functions in any state $|\hat{n}\ran$ are equivalent to those of $\phi^\perp(x,\hat{n})$.  In infinite volume this can be done exactly using projection operators onto different superselection sectors, while in finite volume $V$ we can do it to leading order in $1/V$ by defining a ``$\hat{n}$ operator''
\be
\hat{n}_{op}\equiv \frac{\sqrt{g}}{|m|V}\int  d^3 x \, \phi_i(x)
\ee 
and  replacing $\hat{n}\to \hat{n}_{op}$ in the definition \eqref{phistatedep}.  Since $\lan \hat{n}|\hat{n}_{op}|\hat{n}\ran=\hat{n}$, and the commutator of $\hat{n}_{op}$ with any local operator is another local operator times a power of $1/V$, this operator has the same expectation values as $\phi^\perp_i(x,\hat{n})$ up to $O(1/V)$ corrections.

I claim however that this procedure of ``removing'' the state-dependence does \textit{not} describe what we usually do in the laboratory when studying Goldstone bosons.  Setting up an apparatus to measure this operator would be rather irritating, since it would have to involve the nonlocal operator $\hat{n}_{op}$ each time we measure the field.  What we usually do instead is measure $\hat{n}_{op}$ \textit{once} to determine what state we are in, and then conditioned on the result of this measurement we then measure various combinations of the $\phi^\perp_i(x,\hat{n})$'s as defined in equation \ref{phistatedep}.  The small commutator of the ``order parameter'' $\hat{n}_{op}$ with local operators ensures us that we do not need to measure it again later in the experiment.  The distinction between this protocol and one where we measure the ``state-independent'' operator defined in the previous paragraph by the replacement $\hat{n}\to\hat{n}_{op}$ is not academic; if we started the system in a superposition of different $|\hat{n}\ran$ states, the results would differ substantially (they also differ at any case at $O(1/V)$).  It should be clear however that either is perfectly normal in principle, and they had better both be consistent with quantum mechanics.  I now discuss this more abstractly in the context of general quantum measurement theory.

\subsection{Measurement Theory}
The basic idea of quantum measurement theory is as follows.\footnote{For a nice review see section 3.1 of \cite{preskillnotes}.}  Say we have a system $S$ and we'd like to measure some hermitian operator $A$ that acts on it.  We adjoin the system to a \textit{pointer} system $P$  whose dimensionality is equal to the number of distinct eigenvalues of $A$.  We then arrange for the unitary evolution of the joint system to be
\be\label{measev}
|i\ran_S|0\ran_P\to |i\ran_S |a_i\ran_P,  
\ee
where $|0\ran_P$ is some particular ``initial'' state of the pointer and $|i\ran_S$ is any eigenstate of $A$ with eigenvalue $a_i$.  Note that the states $|a_i\ran_P$ are not necessarily all distinct; different $|i\ran_S$'s could have the same eigenvalues.  If we now start the system in an arbitrary pure state $|\psi\ran_S=\sum_i C_i |i\ran_S$ then we have the evolution
\be
|\psi\ran_S|0\ran_P\to \sum_i C_i |i\ran_S |a_i\ran_P.
\ee
The pointer is now in a mixed state 
\be
\rho_P=\sum_a \sum_{i \, |\, a_i=a}|C_i|^2 |a\ran\lan a|,
\ee
so if we look at it then we will see a result $a$ drawn from the probability distribution predicted by the usual Born rule for measuring $A$.  Of course in this last step we again have to make a measurement, but the pointer is usually assumed to be sufficiently classical that it is ``obvious'' what it means to measure it.

The important point here is that the measurement process can be described as unitary evolution of the system coupled to some apparatus.  The same is true for the protocol described at the end of the previous section, where we first measured $\hat{n}_{op}$ and then conditioned on the result measured some combination of $\phi^\perp_i(x,\hat{n})$'s, but to see it we first need to recall the standard idea of conditioned unitary evolution.  Consider a bipartite system consisting of systems $S_1$ and $S_2$.  We can then define an evolution
\be\label{condun}
|i\ran_{S_1} |j\ran_{S_2}\to |i\ran_{S_1} U_i|j\ran_{S_2},
\ee
which we can interpret as looking at $S_1$ in the basis $|i\ran$ and then, depending on the result, applying a unitary transformation $U_i$ to system $S_2$.  The evolution \eqref{condun} is unitary for any choice of the $U_i$'s.  

\begin{figure}
\begin{center}
\includegraphics[height=5cm]{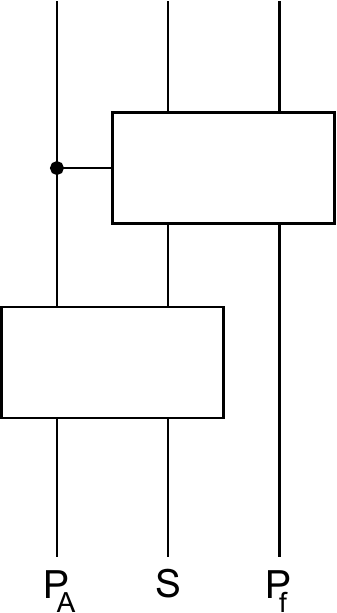}
\caption{The unitary process for measuring $A$ and then conditionally measuring $f$.  Time goes up.  }\label{2meas}
\end{center}
\end{figure}

With these tools we can now give a more general discussion of the two measurement protocols of the previous section.  Say that we have a system $S$ on which $A$ and $B$ are two hermitian operators.  Moreover say that we have some classical function $f(a,b)$ of their eigenvalues.  The reader should think of $A$ as being analogous to $\hat{n}_{op}$ in the previous section.  The first protocol, where we replaced $\hat{n}\to\hat{n}_{op}$ in \eqref{phistatedep}, corresponds to measuring the quantum operator $f(A,B)$ using \eqref{measev}.  Our second protocol, measuring $\hat{n}_{op}$ and then conditionally measuring $\phi^\perp(x,\hat{n})$, generalizes to first to measuring $A$ and finding some result $a$, then measuring the quantum operator $f(a,B)$.  We can describe this as unitary evolution as follows; first adjoin to the system $S$ two pointers, $P_A$ for the first measurement and $P_f$ for the second.\footnote{For simplicity we assume that the number of distinct eigenvalues of $f(a,B)$ is the same for all $a$, enabling us to use just one pointer for the second measurement.}  Then apply the unitary measurement protocol \eqref{measev} to the system $S$ and the first pointer $P_A$.  Finally apply the conditioned unitary evolution \eqref{condun}, where conditioned on the state of $P_A$ we apply the unitary measurement protocol to $S$ and $P_f$ for measuring $f(a,B)$.  The quantum circuit diagram for this evolution is given in figure \ref{2meas},
explicitly the full evolution is
\be
|i\ran_S|0\ran_{P_A}|0\ran_{P_f}\to |i\ran_S|a_i\ran_{P_A}|0\ran_{P_f}\to \sum_{j'}C^i_{j'}|j'\ran_S|a_i\ran_{P_A}|f(a_i,b_{j'})\ran_{P_f},
\ee
where $|i\ran_S$ is an eigenstate of $A$ with eigenvalue $a_i$, $|j'\ran_S$ is an eigenstate of $B$ of eigenvalue $b_{j'}$, and $|i\ran_S=\sum_{j'}C^i_{j'}|j'\ran_S$.  A classical observer can then look at the pointers to sample from the joint distribution for $a$ and $f$ (or the conditional distribution for $f$ given $a$).

I believe that the second protocol captures the essence of what most people think of as ``state-dependent operators'' in ordinary quantum mechanics.  There is some approximately classical observable which we first pin down with a measurement, and then use to decide which other operators to measure.  The entire process can be described as unitary evolution on the system together with an apparatus.

\subsection{State-Dependence in the PR proposal}\label{sdsec3}
I now compare the state-dependence of the PR proposal (or its less-precise earlier cousins) to the above protocols.  To warm up, let's first consider the operators $\mO$ used in building fields \textit{outside} of the horizon.  These apparently depend on some basic properties of the state, for example the mass of the black hole and where it is, but this information is essentially classical.  We are thus in the situation where we can use either of the protocols of the previous subsections to interpret them.  An infalling observer will probably use the second protocol; she will look to see where the black hole is and how big it is before aiming her jump.

The situation for the interior operators $\tO$ is more interesting.  Consider a complete basis of equilibrium states.  We can define some operator $A$ which distinguishes them, and then try to use this information to define state-dependent operators $\tO_a$ for modes behind the horizon.  To run the second protocol we would first measure $A$ and then do a conditional measurement of $\tO_a$.  This would require the infalling observer to do an extremely sensitive measurement of the black hole, essentially determining which microstate it is in.  It is unreasonable to require the infalling observer do this, so we conclude that the second protocol cannot be used to legitimize the PR proposal.  We could also try the first protocol by defining the interior operators including explicitly the operator $A$ in our expressions, which I will denote as $\tO_A$.  We now run into the issue however that the commutator of $A$ with $\mO$ and $\tO_a$ will be quite large.  This then will destroy the algebraic properties of the $\tO_A$'s, and their correlation functions will no longer agree with effective field theory.  

Thus the state-dependence of the PR proposal cannot be interpreted as arising from either of the two protocols we just discussed.  In fact we can go further and argue that there is no possible unitary evolution on the system together with some apparatus which realizes the PR proposal.  More explicitly, there is no single unitary operator which takes an arbitrary equilibrium state together with a given pointer, not depending on the equilibrium state, and uses that pointer to measure the $\tO$ appropriate for the equilibrium state.  To get started I first observe in general that it is impossible to have a pointer which measures two distinct operators: say that $\tO_1$ and $\tO_2$ are two operators associated with different equilibrium states.  Since they are supposed to have the same physical interpretations, they should have the same eigenvalues.  An obvious way to try to get them to both be measured by the same pointer is to find a unitary which implements
\begin{align}\nonumber
|i,1\ran |0\ran&\to |i,1\ran |\tilde{o}_i\ran\\
|i,2\ran |0\ran&\to |i,2\ran |\tilde{o}_i\ran.\label{prevol}
\end{align}
Here $|i,1\ran$ is some complete eigenbasis of $\tO_1$, with eigenvalues $\tilde{o}_i$, and similarly $|i,2\ran$ for $\tO_2$.  It is fairly straightforward to show however that this evolution is only possible if the two operators are in fact equal; for convenience of the reader I give a proof in appendix \ref{proofapp}.\footnote{One might think that the pointer should also be state-dependent, since it is behind the horizon as well, but for simplicity we can take it to be made out of the infalling purple modes in figure \ref{adsinfall}, which are expected to be state-dependent only in the weak sense of the previous two sections.}  The basic idea of the proof is that the first line of \eqref{prevol} completely specifies the unitary operator, so there is no freedom left to fit the second line.  This might be called a ``no state-dependent operators theorem'' of quantum mechanics.  

This theorem does not quite directly address the PR proposal however, since the types of states PR are interested in are equilibrium states and small perturbations thereof, rather than eigenstates of the $\tO$'s.  The same intuition should still apply however; we can introduce a complete basis of equilibrium states, on which the action of the unitary coupling the pointer to the system is fixed.  There would be no remaining freedom to deal with other equilibrium states that are superpositions in this basis.  In fact we can get into trouble even faster by using the observation of section \ref{nonusec} above.\footnote{The argument that follows here is closely related to the ``frozen vacuum'' argument of \cite{Bousso:2013ifa}, but it is reworked a bit to more directly apply to the PR construction.}  For convenience I will work in a simplified version of the qubit evaporation model of section \ref{evapsec}, where I take $B$ to have only a single qubit, I combine $H$ and $R$ into $\bar{B}$, and I take the algebra $\A$ to consist only of operators on $B$.  In any equilibrium state $|\psi\ran$ the density matrix on $B$ will be maximally mixed, so by the Schmidt decomposition we can write
\be
|\psi\ran=\frac{1}{\sqrt{2}}\left(|0\ran_B|\bar{0}\ran_{\bar{B}}+|1\ran_B|\bar{1}\ran_{\bar{B}}\right),
\ee
where $|\bar{0}\ran_{\bar{B}}$ and $|\bar{1}\ran_{\bar{B}}$ are pure states of unit norm that are typically very complicated.  Now let's consider a measurement of $\wt{Z}$, the mirror operator to the $Z$ operator on $B$.  Our ``target'' bulk state is $(|00\ran+|11\ran)/\sqrt{2}$, so by construction measuring $\tilde{Z}$ should produce the same state as measuring $Z$.  We will thus have
\be
|\psi\ran|0\ran_P\to \frac{1}{\sqrt{2}}\left(|0\ran_B|\bar{0}\ran_{\bar{B}}|0\ran_P+|1\ran_B|\bar{1}\ran_{\bar{B}}|1\ran_P\right).
\ee
Let's now consider however the set of four mutually orthogonal equilibrium states:\footnote{These are the types of equilibrium states one would construct acting on $|\psi\ran$ with the ``unitary behind the horizon'' type operators discussed in section \ref{nonusec}, but here I will follow the rules of PR and construct mirror operators which see the ``horizon'' as unexcited.}
\begin{align}\nonumber
|\psi_\pm\ran&=\frac{1}{\sqrt{2}}\left(|0\ran_B|\bar{0}\ran_{\bar{B}}\pm|1\ran_B|\bar{1}\ran_{\bar{B}}\right)\\
|\chi_\pm\ran&=\frac{1}{\sqrt{2}}\left(|0\ran_B|\bar{1}\ran_{\bar{B}}\pm|1\ran_B|\bar{0}\ran_{\bar{B}}\right).
\end{align}
Since these are all equilibrium states, measuring $\tilde{Z}$ should in all cases be equivalent to measuring $Z$ so after taking superpositions we have the following evolution
\begin{align}\nonumber
|0\bar{0}\ran_{B\bar{B}}|0\ran_P&\to |0\bar{0}\ran_{B\bar{B}}|0\ran_P\\\nonumber
|1\bar{1}\ran_{B\bar{B}}|0\ran_P&\to |1\bar{1}\ran_{B\bar{B}}|1\ran_P\\\nonumber
|0\bar{1}\ran_{B\bar{B}}|0\ran_P&\to |0\bar{1}\ran_{B\bar{B}}|0\ran_P\\
|1\bar{0}\ran_{B\bar{B}}|0\ran_P&\to |1\bar{0}\ran_{B\bar{B}}|1\ran_P.
\end{align}
So far this evolution can be unitary, since after all it is equivalent to measuring $Z$, which is state-independent.  It cannot however be unitary if it is restricted to act only on $P$ and $\bar{B}$, which is after all what we should demand; the infalling observer knows that she is measuring $\wt{Z}$, not $Z$, since these are done by different physical experiments.  Indeed this would require both $|\bar{0}0\ran_{\bar{B}P}\to |\bar{0}0\ran_{\bar{B}P}$ and $|\bar{0}0\ran_{\bar{B}P}\to |\bar{0}1\ran_{\bar{B}P}$, as well as both $|\bar{1}0\ran_{\bar{B}P}\to |\bar{1}0\ran_{\bar{B}P}$ and $|\bar{1}0\ran_{\bar{B}P}\to |\bar{1}1\ran_{\bar{B}P}$.  More operationally we could instead demand that measuring $\wt{Z}$, applying $X$ to flip the qubit $B$, and then measuring $\wt{Z}$ again returns the same result for both $\wt{Z}$ measurements, which leads to a similar contradiction.  By presenting the argument this way we see that this contradiction is closely related to the ambiguity of section \ref{nonusec}; acting with $X$ sends $|\psi_+\ran\to|\chi_+\ran$, so if the formalism itself cannot decide whether or not $|\chi_+\ran$ is excited we can hardly expect a pointer to be able to.  

\section{Including the Infalling Observer}\label{obsec}
We have now seen that the PR proposal is inconsistent with quantum mechanics on two serious counts. We saw in section \ref{nonusec} that it can assign to a single quantum state more than one physical interpretation, and we saw in section \ref{sdsec} that its measurement process cannot be realized as unitary evolution, as opposed to ordinary quantum mechanics (with pointers) where it can.  What then are we to conclude?  It seems that what the proposal needs to work is some rule along the lines of the following: say I am going to jump into a black hole, which is in some equilibrium state $|\psi\ran$.  Moreover say that you are planning to act on the state with some operator $\wt{U}$ (or $V$) as defined in section \ref{nonusec}.  If I \textit{know} that you are going to do this, then I conclude that I will see an excited horizon, and if I jump in that is what I see.  But if I \textit{don't} know you are going to do this, then my $\tO$ operators are automatically redefined in such a way that I see a smooth horizon when I jump in, even though the quantum state of the black hole is the same in either case.  The full state of the system in the two different situations wouldn't actually be the same, since the internal state of the observer is different in the two cases.  Is this crazy?  Undoubtedly, but that does not automatically mean that it is wrong.\footnote{In fact it is somewhat reminiscent of the Gottesman-Preskill refinement of the black hole final state proposal, where any thing happening behind the horizon is ``unhappened'' by post-selection from the point of view of somebody outside the horizon but presumably not for somebody who falls in \cite{Horowitz:2003he,Gottesman:2003up,Lloyd:2013bza,Bousso:2013uka}.  It would be interesting to understand better the relationship between that proposal and the one under consideration here.}  Rather than prolonging this paper further by trying to make a consistent theory that accommodates this kind of thing, which is what I expect would really be needed to make some version of the PR proposal (or $A=R_B$, ER=EPR, etc) work for black holes in generic states, I will instead close by making some general comments about the validity of quantum mechanics for infalling observers.\footnote{There seems to be some overlap with some of the ideas here and what Mathur and Turton call ``fuzzball complementarity'' \cite{Mathur:2013gua}, although I disagree with various parts of their discussion.}  

In ordinary situations where we study quantum mechanics, the system under study is ``small'' and our apparatus is ``big''.  This allows us to basically treat the apparatus classically, up to a single pointer variable as we have discussed in the previous section.  The detailed history and construction of the apparatus (and the experimenter) are completely irrelevant for the outcome of the experiment, at least as long as the experiment has been constructed correctly.  This is to be contrasted with the type of thing described in the previous paragraph, where what the experimenter is aware of and intends to do is of paramount importance.  It is interesting to note however that in the black hole situation the ``small/big'' situation is reversed; we are trying to study a giant black hole as a quantum mechanical object, using an infalling observer who is rather small in comparison.  This is unlike any other situation where we have tested quantum mechanics, and it does not seem a priori absurd to imagine that the usual measurement theory would need to be modified in this case.\footnote{In the context of cosmology it has already been argued that the experiences of observers with fundamentally limited resources do not have precise quantum mechanical descriptions \cite{Harlow:2010my,Bousso:2011up}, and any lesson of this nature which we could learn from black holes would obviously be very valuable for cosmology.}  The infalling observer simply cannot carry in an apparatus which is able to record any substantial fraction of the information which would be needed to describe the black hole in detail, since this would by necessity cause large back-reaction.\footnote{This can be quantified using the recently proven Bekenstein-Casini bound \cite{Bekenstein:1980jp,Casini:2008cr}, which roughly says that an object capable of storing $S$ bits of information much have a mass greater than $S/R$, where $R$ is the size of the object. To fit in the black hole the object must be small compared to the black hole, while to avoid back-reaction its mass must be small compared to the black hole mass; taken together with the bound these show that the memory capacity $S$ of the object must be much smaller than the black hole entropy.}  AMPS have tried to avoid this problem by inventing an experiment that does not require the infalling observer to actually carry in a large number of qubits, but they do need to at some point perform a complicated experiment on an $O(1)$ fraction of the Hawking radiation.  The original AMPS experiment involved an extremely sophisticated quantum computation which is probably impossible to really implement \cite{Harlow:2013tf}, but as we saw in section \ref{evapsec}, for the PR construction to fail one only needs to consider something like flying around and flipping the helicity of some $O(1)$ fraction of the Hawking photons.  This is much easier than the AMPS quantum computation, there is no principled reason why it cannot be done.  The point here however is that although the infalling observer can remember that a large number of photon helicities were flipped, she cannot actually carry in a list of which photons were flipped and which weren't.  We might imagine that the definition of the $\tO$'s gets reset in this case and she sees a smooth horizon, with her inability to actually remember (or carry in a record of) which ones were flipped preventing various paradoxes.  Should the theory really make use of this type of thing?  One might hope not, but if the PR proposal or something like it is to prevent firewalls in generic states it seems more and more likely that it will have to.  A solid example of a theory that does this (or a concrete explanation of how AdS/CFT secretly does it) would obviously be necessary before it could really be taken seriously.   

\paragraph{Acknowledgments} I'd like to thank Herman Verlinde, Edward Witten, and especially Juan Maldacena, Kyriakos Papadodimas,and Suvrat Raju for many helpful discussions of the PR proposal.  I'd also like to thank Raphael Bousso, Xi Dong, Don Marolf, Shiraz Minwalla, Joe Polchinski, Vladimir Rosenhaus, Douglas Stanford, Steve Shenker, Jaimie Sully, Leonard Susskind, and Erik Verlinde for useful conversations.  I am supported by the Princeton Center for Theoretical Science.

\appendix
\section{Gauge Constraints and Charges}\label{gaugeapp}
In theories with gauge symmetries it is not necessarily the case that a decomposition of space into two regions should induce a tensor factorization of the Hilbert space.  The basic reason is that the constraints used in defining the physical Hilbert space typically involve spatial derivatives, for example the Gauss law constraint 
\be
\nabla \cdot E=\rho
\ee
in electrodynamics or the Hamiltonian constraint
\be
K_{ij}K^{ij}-K^2+R^{\{3\}}=16\pi G T_{00}
\ee
in gravity.  Since the PR construction of mirror operators is based on assuming a tensor factorization of the bulk Hilbert space as in equation \eqref{psib}, we need to make sure that this is not inconsistent with the constraint structure of the bulk gauge symmetries.  

Similarly in defining conserved charges associated to gauge symmetries, the constraints typically imply that the charges can be written as surface integrals over the boundary of the spatial region in question.  In the PR construction we are replacing two asymptotic boundaries with one, and we need to make sure that the relationship between the CFT Hamiltonian and the bulk ADM Hamiltonians is chosen consistently.  

The purpose of this appendix is to address these concerns in more detail, motivated by a simple example which illustrates the relevant issues. Readers who are willing to accept the tensor factorization and are only interested in understanding my choice of the commutator of $H$ and the $\tO$'s being zero can perhaps skip to subsection \eqref{hamsec}. Other work that discusses some of the issues in this appendix includes \cite{Donnelly:2012st,Casini:2013rba,Radicevic:2014kqa,Heemskerk:2012np,Papadodimas:2013jku}. 

\subsection{A Toy Model}
The constraint structure of perturbative gravity in the AdS-Schwarzschild background is a bit complicated, but an excellent model which captures the relevant physics is scalar electrodynamics in $1+1$ dimensions, quantized on a spatial interval of finite length.  I will take the boundary conditions to be $A_0=0$ and $\phi=0$ at each endpoint, and require that gauge transformations vanish there.\footnote{This system is the $1+1$ dimensional analogue of a region of space between two perfect conductors that are connected by a wire.  These boundary conditions are chosen because they resemble the usual ``normalizeable'' boundary conditions in AdS that lead to well-defined asymptotic charges.}  The Hamiltonian is
\be\label{qedH}
H=\int_0^1 dx \left[\frac{1}{2}\Pi^2+\pi^\dagger \pi+\left(D_x\phi\right)^\dagger D_x\phi\right],
\ee
where
\be
D_\mu \phi=\partial_\mu\phi-iqA_\mu \phi,
\ee 
and
\be
\Pi\equiv -E= \partial_x A^0+\dot{A}_x
\ee
is the conjugate momentum to $A_x$.
\be
\pi=(D_0\phi)^\dagger
\ee
is the conjugate momentum to $\phi$.  Physical states obey the Gauss law constraint
\be
\left(\partial_x \Pi+\rho\right)|\psi\ran=0,
\ee
where
\be
\rho=iq\left(\phi \pi-\pi^\dagger \phi^\dagger\right)
\ee
is the charge density.

An interesting property of these boundary conditions is that they allow us to define nonlocal gauge-invariant operators by ``dressing'' charged fields with Wilson lines.  For example the operators
\begin{align}\nonumber
\overrightarrow{\phi}(x)&\equiv \phi(x)e^{iq\int_x^1 dx A_x}\\\nonumber
\overleftarrow{\phi}(x)&\equiv \phi(x)e^{-iq \int_0^x dx A_x}\\\label{invops}
W&\equiv e^{iq\int_0^1 dx A_x}
\end{align}
are gauge invariant, with the first operator being the product of the second and third.

Another interesting property of these boundary conditions is that they do not allow us to use ``axial gauge'' $A_x=0$; for example $\phi=0$, $A_0=0$, and $A_x=E_0 t$ is a nontrivial solution of the equations of motion that obeys the boundary conditions but cannot be put into axial gauge without violating them.  

For maximal clarity it is convenient to put this model on a spatial lattice, which I will take to have four points.  I will work in units where the lattice spacing is one.  For my choice of boundary condition there are three gauge field degrees of freedom living on the links between the points, $A_{12}$, $A_{23}$, and $A_{34}$, and there are two charged fields living on the middle two lattice sites, $\phi_2$ and $\phi_3$.  The gauge group is $U(1)\times U(1)$, with the gauge transformations living at sites $2$ and $3$.  Since charge fields have $2$ degrees of freedom each, there are a total of $7$ non-gauge-invariant degrees of freedom.  There are thus $5$ physical degrees of freedom; more concretely we can write any gauge-invariant wave function as
\be\label{invwf}
\lan A_{12}A_{23}A_{34}\phi_2\phi_2^{*}\phi_3\phi_3^*|\psi\ran=\psi\left[W,\overleftarrow{\phi}_2,\overleftarrow{\phi}_2^*,\overrightarrow{\phi}_3,\overrightarrow{\phi}_3^*\right].
\ee
Here we have
\begin{align}\nonumber
\overrightarrow{\phi}_3&= \phi_3e^{iqA_{34}}\\\nonumber
\overleftarrow{\phi}_2&= \phi_2e^{-iq A_{12}}\\\label{latinvops}
W&=e^{iq(A_{12}+A_{23}+A_{34})}.
\end{align}
There are also gauge invariant canonical momentum operators
\begin{align}\nonumber
\overrightarrow{\pi}_3&=\pi_3e^{-iqA_{34}}\\
\overleftarrow{\pi}_2&=\pi_2e^{iqA_{12}}.
\end{align}
The Gauss law constraint becomes
\begin{align}\label{latgcons}
E_{34}-E_{23}&=\rho_3\\
E_{23}-E_{12}&=\rho_2.
\end{align}
If we wish to go to axial gauge, the most we can do is for example to set $A_{12}=A_{34}=0$; as expected from the continuum argument above there is not enough gauge symmetry to also set $A_{23}=0$.  The Wilson line $W$ is thus a physical degree of freedom, it cannot be removed by gauge fixing.  This choice of gauge does remove the manifestly nonlocal dressing from $\overrightarrow{\phi}_3$ and $\overleftarrow{\phi}_2$, but there will still be a nonlocal commutator with the electric field that remembers it (it doesn't look particularly nonlocal here, but in the obvious generalization to more lattice sites it will). 

This discussion of the physical Hilbert space makes it clear that if we cut the interval in half, there is no gauge-invariant tensor factorization of the Hilbert space associated with this.  In addition to the charged operators acting on sites $2$ and $3$, there is the Wilson line that cannot be generated by gauge invariant operators localized on one side or the other (this again is a bit more obvious if we include more lattice points).\footnote{As explained in \cite{Casini:2013rba}, this inability to factorize will be the case anytime the subalgebra of operators associated with a subregion has a nontrivial center.  In this model we should clearly include $\overleftarrow{\phi}_2$ and its canonical conjugate $\overleftarrow{\pi}_2$ in the subalgebra associated to the ``left half'' of the interval, as well as $E_{12}$.  It is up to us whether or not we include $E_{23}$; if we do then $E_{23}$ is a nontrivial element of the center, since acting on gauge invariant states it is equal to $E_{34}-\rho_3$, which obviously commutes with everything else in the subalgebra.  If we do not include $E_{23}$, then $E_{12}+\rho_2$ is a nontrivial element of the center.} Since a tensor factorization is necessary for the PR construction, it seems we are in a bit of trouble.  More explicitly, acting on a generic state of the form \eqref{invwf} with an operator $f(\overleftarrow{\phi}_2,\overleftarrow{\pi}_2,E_{12})$ is not in general equal to the action of some other operator $g(\overrightarrow{\phi}_3,\overrightarrow{\pi}_3,E_{34})$ on the same state.  It \textit{will} typically be equal to the action of some operator $g(\overrightarrow{\phi}_3,\overrightarrow{\pi}_3,E_{34},W)$, but this depends also on $W$, which in the gravitational analogue has no single-CFT representation.\footnote{One might be tempted to try to mirror the ``left'' degrees of freedom \textit{and} the Wilson line $W$ into the ``right'' degrees of freedom, but this is in general not possible since the matrix $C$ in \eqref{psib} can only be invertible in the relevant sense if the set of degrees of freedom we are ``mirroring to'', ie the ``right'' degrees of freedom, is not smaller than the set of degrees of freedom we are mirroring from.}  This dependence will go away only if for some reason we are only interested in states where the Wilson line degree of freedom factors out, that is in states whose wave functions have the form\footnote{We can also consider states where the tensor factorization is in terms of $\overrightarrow{\phi}_3$ and $\overrightarrow{\phi}_2$.  The discussion below would be similar, although some words would change, but this choice is a better analogy for what we usually do in the two-sided bulk system.}
\be\label{factorwf}
\psi\left[W,\overleftarrow{\phi}_2,\overleftarrow{\phi}_2^*,\overrightarrow{\phi}_3,\overrightarrow{\phi}_3^*\right]=\psi_W[W] \psi_\phi\left[\overleftarrow{\phi}_2,\overleftarrow{\phi}_2^*,\overrightarrow{\phi}_3,\overrightarrow{\phi}_3^*\right].
\ee

There is indeed a fairly natural set of mutually orthogonal subspaces that have this form: the subspaces where $E_{23}$ is a constant.  These subspaces have the nice property that they are preserved by the action of $E_{12}$, $E_{34}$, $\overleftarrow{\phi}_2$, $\overrightarrow{\phi}_3$, their hermitian conjugates, and their conjugate momenta.  They also relate nicely to the definition of charge; the total charge is given by a sum of two ``boundary terms''
\be
Q_{tot}\equiv\rho_2+\rho_3=E_{34}-E_{12},
\ee
and we can also define ``left'' and ``right'' charge operators
\begin{align}\nonumber
\hat{Q}_{R}&=E_{34}-E_{23}=\rho_3\\
\hat{Q}_L&=E_{23}-E_{12}=\rho_2,\label{QRL}.
\end{align}
On a subspace of constant $E_{23}$ we have the nice property that the boundary operators $E_{12}$ and $E_{34}$ act only on the left and right tensor factors respectively.  A crucial choice in the PR construction is what combination of these operators we interpret as the charge operator (or Hamiltonian for the case of gravity) when we have only a single CFT.   

\subsection{The Two-sided Gravitational Bulk}
I now more heuristically discuss the case of gravity in the two-sided asymptotically-AdS system.  We'd like to split the system into two parts, but as in electrodynamics the constraints do not allow us to do this.  There is no simple gravitational analogue of Wilson lines, so the construction of gauge invariant operators is more complicated.  I will leave the details to future work, and just make a few comments about what I expect to happen based on the example just discussed.  

In order to effectively get a tensor product Hilbert space for the purposes of the PR construction, we again need to impose some sort of additional constraint analogous to $E_{23}$ being a constant.  In Einstein gravity an appealing gauge-invariant proposal is that we should demand that the area of the extremal-area bulk surface $\Sigma$ of topology $S^{d-1}$ be given by a c-number acting on the state.  In more general theories (which we will need to consider if we want to get the $1/N$ corrections right) we might instead ask that the integral of the Noether charge $d-1$ form $Q[\xi]$ \cite{Wald:1993nt,Iyer:1994ys} over the bulk surface $\Sigma$ of topology $S^{d-1}$ that extremizes it be constant.  Here $\xi$ is any timelike vector field which approaches the asymptotic time translation vector $\partial_t$ at the right boundary, which approaches $-\partial_t$ at the left boundary, and which vanishes on $\Sigma$ and acts in the vicinity of $\Sigma$ as the boost generator in the two-dimensional plane orthogonal to $\Sigma$.  This determines $\xi$ only up to a constant multiple, so $Q$ defined this way has an irrelevant normalization ambiguity.  In Einstein gravity 
\be\int_\Sigma Q[\xi]\propto \frac{A}{4G}.
\ee

One motivation for this proposal is that in any gravity theory the canonical Hamiltonian that evolves the metric and matter fields forwards along $\xi$ to the right of $\Sigma$, while keeping them fixed to the left, is \cite{Iyer:1994ys} 
\be
\hat{H}_R=H_R-\int_\Sigma Q,
\ee 
where $H_R$ is the AdS version of the ADM Hamiltonian, perhaps including $1/N$ corrections to Einstein gravity, which is a boundary integral at the right boundary.\footnote{I thank Don Marolf for emphasizing the existence and possible importance of this boundary term on several occasions.}  $\hat{H}_R$ here is analogous to $\hat{Q}_R$ in equation \eqref{QRL}, and there we saw that we needed to set $E_{23}$ to a constant in order to have $\hat{Q}_R$ be given only by a boundary term at infinity.  Another motivation is that any construction of gauge-invariant matter operators in the bulk that proceeds by evolving operators in region I in from the right boundary and operators in region III in from the left boundary should always produce operators that commute with $\int_\Sigma Q$, since their ``gravitational dressing'' will always extend away from $\Sigma$.  In Einstein gravity this point is supported by the calculations of Shenker and Stanford \cite{Shenker:2013yza}, who saw that in a wide variety of states produced by acting with local operators on one or both sides, the area of the extremal surface is never modified.  Even more generally a recent theorem \cite{headrick} shows that the extremal surface always lies in the ``causal shadow'' of the two boundaries, meaning that it can never receive or send signals from either boundary.  It seems quite plausible that the subspace of states where this quantity is fixed factorizes into left and right degrees of freedom that provide a reasonable ``laboratory'' for the PR construction of the interior.\footnote{One subtlety here is that in two-sided states where the extremal surface is not also a bifurcate horizon, it is probably not true the bulk operators with simple two-CFT prescriptions are really sufficient to give initial conditions for the bulk evolution up to the region behind the horizon done in the PR construction.  This is related to the problems with CFT states of narrow energy width described in section \ref{TFDsec}. Another important question is whether or not the TFD state actually has the property that $\int_\Sigma Q$ doesn't fluctuate.}

Independent of the validity of the PR construction, it is interesting to contemplate the meaning of bulk states that are superpositions of different $\int_\Sigma Q$ in the context of AdS/CFT.  Equivalently, if there is a bulk gauge field it is interesting to contemplate the meaning of states where the electric flux through $\Sigma$ is not definite.  Does an operator representing a bulk Wilson line extending from one boundary to the other exist on the Hilbert space of the two CFT's?  It is far from clear that it does, since after all the Hilbert space of the two CFT's trivially factorizes into left and right parts, each of which is separately gauge-invariant, and as we saw above the Wilson line does not respect this factorization.  What are we to make of this?  One option is to argue that this means that the two CFT's do not give a complete description of the bulk physics; we need to include another degree of freedom to describe this Wilson line.  This is rather similar to the ``superselection sectors'' of Marolf and Wall \cite{Marolf:2012xe}, although we seem to have arrived at it from a rather different direction here.  It is also reminiscent of what is sometimes called ``strong complementarity'' \cite{harlow2,Bousso:2012as,Harlow:2013tf}.  Alternatively it may be that it is the CFT's which are correct, and that the bulk observer has deluded herself into thinking that this Wilson line operator should exist.  One encouraging point is that there \textit{are} states in the two CFT's which we interpret as having some electric flux through $\Sigma$; the Reissner-Nordstrom wormholes.  We cannot assemble them in a simple way from an uncharged TFD state by throwing in charges from the two sides, even if we do it in a correlated manner, but they do exist.  Does this mean that the bulk Wilson line must also exist as an operator in the CFT's in some nontrivial way?  Since the bulk gauge field (and bulk graviton) are ``emergent'' in the sense of not really being present in the fundamental CFT description of the theory, perhaps the UV regularization provided by quantum gravity is smart enough to avoid the difficulties of factorizing present in the lattice model.  The description of this Wilson line seems like a potentially valuable toy version of the description of the interior in general, and I hope to have more to say about it soon. 

\subsection{What is the Hamiltonian?}\label{hamsec}
The PR construction works by finding a set of single-CFT operators whose algebra and action on equilibrium states reproduces that of the two-sided bulk theory.  One question that is not entirely clear is which two-sided bulk operator should be represented by the CFT Hamiltonian $H$.  There are two somewhat natural candidates: 
\begin{align}\nonumber
H&\to H_R-H_L+E_0\\
H&\to H_R.
\end{align}
Since we are in the sector where $\int_\Sigma Q$ is a constant, $H_R$ acts on the right-side factor of the Hilbert space and will commute with any operator in region III; if we represent it with the CFT Hamiltonian $H$ then our mirror operators $\tO$ should be taken to commute with $H$.  This is the choice I have made in the main text.  Alternatively if we take $H_R-H_L+E_0$ to be represented as $H$, then the mirror operators should not commute with $H$;\footnote{Here $E_0$ is a c-number that is included so that the bulk expectation value is consistent with the expected energy in the CFT.} we instead have
\begin{align}\nonumber
[H,\tO_\omega]\mathcal{H}_\psi&=\omega \tO_\omega \mathcal{H}_\psi\\
[H,\tO_\omega^\dagger]\mathcal{H}_\psi&=-\omega \tO_\omega^\dagger \mathcal{H}_\psi.\label{PRcom}
\end{align}
 This is the choice advocated by Papadodimas and Raju, but I will argue in the remainder of this section that it is problematic.  

The equilibrium condition \eqref{eqcond} will be the same for expectation values of elements of $\A$ built only from $\mO$'s and their $C$ conjugates for either interpretation of $H$, but we should apply it also to expectation values involving $H$ only if we take $H$ to represent $H_R$.  If we take $H$ to represent $H_R-H_L+E_0$ then in the bulk this is not an operator that acts on the right only, so \eqref{eqcond} should not apply to it.  In fact it is fairly straightforward to show that if \eqref{eqcond} does apply to elements of $\A$ that include the Hamiltonian, then we essentially \textit{must} take $[H,\tO]=0$.  Indeed consider two algebra elements $A_\alpha$ and $A_\beta$ which are made only out of $\mO$'s. By using equation \eqref{mireq}, the fact that $[\tO,\mO]=0$, and the equilibrium condition \eqref{eqcond}, we see that 
\begin{align}\nonumber
\lan\psi|A_\alpha [H,\tO] A_\beta|\psi\ran&=\lan\psi|A_\alpha H A_\beta C \mO^\dagger C^{-1}|\psi\ran-\lan\psi|C^{-1 \dagger}\mO^\dagger C^\dagger A_\alpha H A_\beta|\psi\ran\\\nonumber
&=\tr\left(C^\dagger A_\alpha H A_\beta C\mO^\dagger\right)-\tr\left(C \mO^\dagger C^\dagger A_\alpha H A_\beta\right)\\
&=0.\label{comzero}
\end{align}
This certainly is incompatible with \eqref{PRcom}, if the commutator is isn't zero then it must apparently be proportional to a somewhat strange operator whose expectation value between any states in $\mathcal{H}_\psi$ produced by acting just with $\mO$'s is zero but which is not zero between elements when the Hamiltonian is involved.  

To get an idea of what would be necessary, let's consider what type of CFT states could be compatible with the TFD state if we choose $H$ to represent $H_R-H_L+E_0$.  By consistency with the bulk we must have
\be\label{weird}
CHC^{-1}|\psi\ran=E_0|\psi\ran+O(e^{-cS}),
\ee 
which follows from the bulk equation \eqref{bulkmir} applied to $H_R$:
\be
H_L|\psi_{bulk}\ran=CH_RC^{-1}|\psi_{bulk} \ran.
\ee 
For the TFD $C$ commutes with $H$, so \eqref{weird} says that CFT states that are compatible with the TFD must be energy eigenstates to within exponential precision, at least as far as expectation values of elements of $\A$ are concerned.  This is rather bizarre, since the equilibrium condition \eqref{eqcond} still applies to elements of $\A$ that are only made out of $\mO$'s; apparently we need states which ``fake'' a thermal distribution to all orders in $1/N$ for expectation values involving only $\mO$'s, even though they are actually almost energy eigenstates.  This is surprising from the point of view of the discussion of section \eqref{TFDsec}, where we saw that the difference between microcanonical and canonical expectation values entered at low orders of perturbation theory in $1/N$.  At best we could try to achieve this by detailed microscopic tuning of the state, which is to be compared with the natural set of states \eqref{thermalpure} that are compatible with the TFD state if we take $H$ to represent $H_R$.    

In fact the structure of the operator product expansion in the CFT basically ensures that if the equilibrium condition \eqref{eqcond} applies to operators that are even fairly simple functions of $\mO$, it must also apply to $H$.  In any CFT the conformal Ward identity ensures that the stress tensor $T_{\mu\nu}$ must appear in the $\mO \mO$ OPE, and we can isolate its contribution by subtracting off a few relevant operators which are all built from elements of the algebra $\A$.  Since the Hamiltonian is just the zero mode of $T_{00}$, we can write a ``formula'' for it in terms of the $\mO$'s; we can then run the argument \eqref{comzero} to conclude that we must take $[H,\tO]=0$.\footnote{More conservatively if we include the various restrictions on elements of $\A$ we may only be able to produce the Hamiltonian this way to within $1/N$ corrections in expectation values, but that should be enough to rule out \eqref{PRcom}.}  

For this reason I have taken $H$ to represent $H_R$ in the main text.  This proposal is still not completely satisfactory, for example it has the problem with states with narrow energy width discussed in section \ref{TFDsec}, but at the moment it seems to be the only possibility based on the general strategy of consistently simulating the two-sided bulk order by order in $1/N$ in a single CFT.  One alternative which is perhaps worth exploring further is to instead attempt to directly simulate the one-sided bulk including the collapse in the CFT; of course this is what AdS/CFT does normally, but the new ingredient would be to allow state-dependence and a restricted algebra $\A$ in an attempt to find the red modes behind the horizon in figure \ref{adsinfall}.  Unfortunately doing this would essentially require us to start over from the beginning, with considerably more complications and possible ambiguities, and I won't attempt it here.

\section{A Proof}\label{proofapp}
Here I give the proof of a result quoted in section \ref{sdsec3}; that a given pointer cannot measure two different operators.  The two operators are called $\tO_1$ and $\tO_2$, and they are assumed on physical grounds to have the same eigenvalues.  To begin with I will also assume that the eigenvalues have the same degeneracies, but this will be relaxed in the end.  We want to show that if there exists a unitary which implements 
\begin{align}\nonumber
|i,1\ran |0\ran&\to |i,1\ran |\tilde{o}_i\ran\\
|i,2\ran |0\ran&\to |i,2\ran |\tilde{o}_i\ran,\label{prevola}
\end{align}
where $|i,1\ran$ is some complete eigenbasis of $\tO_1$, with eigenvalues $\tilde{o}_i$, and similarly $|i,2\ran$ for $\tO_2$, then the two operators must be equal.  We can always write
\be\label{21exp}
|i,2\ran=\sum_j C_{ij} |j,1\ran,
\ee
where $C_{ij}$ is some unitary matrix.  The consistency of \eqref{prevola} and \eqref{21exp} requires
\be
\sum_{j \, |\, \tilde{o}_j=\tilde{o}}C_{i j}|j,1\ran=
\begin{cases}
|i,2\ran & \tilde{o}_i=\tilde{o}\\
0 & \tilde{o}_i\neq\tilde{o}
\end{cases}
\ee
for all $\tilde{o}$ and $i$, which then implies that the unitary $C_{ij}$ is block-diagonal on the subspaces of definite $\tO_1$.  If we look at the spectral representations of the operators
\begin{align}
\tO_1&=\sum_i \tilde{o}_i |i,1\ran\lan i,1|\\
\tO_2&=\sum_i \tilde{o}_i |i,2\ran\lan i,2|,
\end{align}
we see immediately that we have shown that
\be
\tO_1=\tO_2,
\ee
so the operators weren't actually state-dependent in the first place.  The extension to the case where we allow the operators to possibly have different multiplicities is straightforward; the same argument shows that $C_{ij}$ is nonzero only if $i$ and $j$ have the same eigenvalues, but this shows that $C$ and $C^\dagger$ map subspaces with the same eigenvalue into each other.  By the unitarity of $C$ this can only be possible if these subspaces have the same dimensionalities.
\bibliographystyle{jhep}
\bibliography{bibliography}

\providecommand{\href}[2]{#2}\begingroup\raggedright\begin{thebibliography}{10}

\bibitem{Hawking:1974sw}
S.~Hawking, {\it {Particle Creation by Black Holes}},  {\em Commun.Math.Phys.}
  {\bf 43} (1975) 199--220.

\bibitem{Hawking:1976ra}
S.~Hawking, {\it {Breakdown of Predictability in Gravitational Collapse}},
  {\em Phys.Rev.} {\bf D14} (1976) 2460--2473.

\bibitem{Banks:1996vh}
T.~Banks, W.~Fischler, S.~Shenker, and L.~Susskind, {\it {M theory as a matrix
  model: A Conjecture}},  {\em Phys.Rev.} {\bf D55} (1997) 5112--5128,
  [\href{http://xxx.lanl.gov/abs/hep-th/9610043}{{\tt hep-th/9610043}}].

\bibitem{Maldacena:1997re}
J.~M. Maldacena, {\it {The Large N limit of superconformal field theories and
  supergravity}},  {\em Adv.Theor.Math.Phys.} {\bf 2} (1998) 231--252,
  [\href{http://xxx.lanl.gov/abs/hep-th/9711200}{{\tt hep-th/9711200}}].

\bibitem{Witten:1998qj}
E.~Witten, {\it {Anti-de Sitter space and holography}},  {\em
  Adv.Theor.Math.Phys.} {\bf 2} (1998) 253--291,
  [\href{http://xxx.lanl.gov/abs/hep-th/9802150}{{\tt hep-th/9802150}}].

\bibitem{Gubser:1998bc}
S.~Gubser, I.~R. Klebanov, and A.~M. Polyakov, {\it {Gauge theory correlators
  from noncritical string theory}},  {\em Phys.Lett.} {\bf B428} (1998)
  105--114, [\href{http://xxx.lanl.gov/abs/hep-th/9802109}{{\tt
  hep-th/9802109}}].

\bibitem{Mathur:2009hf}
S.~D. Mathur, {\it {The Information paradox: A Pedagogical introduction}},
  {\em Class.Quant.Grav.} {\bf 26} (2009) 224001,
  [\href{http://xxx.lanl.gov/abs/0909.1038}{{\tt arXiv:0909.1038}}].

\bibitem{Giddings:2011ks}
S.~B. Giddings, {\it {Models for unitary black hole disintegration}},  {\em
  Phys.Rev.} {\bf D85} (2012) 044038,
  [\href{http://xxx.lanl.gov/abs/1108.2015}{{\tt arXiv:1108.2015}}].

\bibitem{Braunstein:2009my}
S.~L. Braunstein, S.~Pirandola, and K.~Życzkowski, {\it {Better Late than
  Never: Information Retrieval from Black Holes}},  {\em Phys.Rev.Lett.} {\bf
  110} (2013), no.~10 101301, [\href{http://xxx.lanl.gov/abs/0907.1190}{{\tt
  arXiv:0907.1190}}].

\bibitem{Almheiri:2012rt}
A.~Almheiri, D.~Marolf, J.~Polchinski, and J.~Sully, {\it {Black Holes:
  Complementarity or Firewalls?}},  {\em JHEP} {\bf 1302} (2013) 062,
  [\href{http://xxx.lanl.gov/abs/1207.3123}{{\tt arXiv:1207.3123}}].

\bibitem{Marolf:2012xe}
D.~Marolf and A.~C. Wall, {\it {Eternal Black Holes and Superselection in
  AdS/CFT}},  {\em Class.Quant.Grav.} {\bf 30} (2013) 025001,
  [\href{http://xxx.lanl.gov/abs/1210.3590}{{\tt arXiv:1210.3590}}].

\bibitem{Almheiri:2013hfa}
A.~Almheiri, D.~Marolf, J.~Polchinski, D.~Stanford, and J.~Sully, {\it {An
  Apologia for Firewalls}},  {\em JHEP} {\bf 1309} (2013) 018,
  [\href{http://xxx.lanl.gov/abs/1304.6483}{{\tt arXiv:1304.6483}}].

\bibitem{Bousso:2013wia}
R.~Bousso, {\it {Firewalls From Double Purity}},  {\em Phys.Rev.} {\bf D88}
  (2013) 084035, [\href{http://xxx.lanl.gov/abs/1308.2665}{{\tt
  arXiv:1308.2665}}].

\bibitem{Marolf:2013dba}
D.~Marolf and J.~Polchinski, {\it {Gauge/Gravity Duality and the Black Hole
  Interior}},  {\em Phys.Rev.Lett.} {\bf 111} (2013) 171301,
  [\href{http://xxx.lanl.gov/abs/1307.4706}{{\tt arXiv:1307.4706}}].

\bibitem{Papadodimas:2013wnh}
K.~Papadodimas and S.~Raju, {\it {The Black Hole Interior in AdS/CFT and the
  Information Paradox}},  \href{http://xxx.lanl.gov/abs/1310.6334}{{\tt
  arXiv:1310.6334}}.

\bibitem{Papadodimas:2013jku}
K.~Papadodimas and S.~Raju, {\it {State-Dependent Bulk-Boundary Maps and Black
  Hole Complementarity}},  \href{http://xxx.lanl.gov/abs/1310.6335}{{\tt
  arXiv:1310.6335}}.

\bibitem{Bousso:2012as}
R.~Bousso, {\it {Complementarity Is Not Enough}},  {\em Phys.Rev.} {\bf D87}
  (2013) 124023, [\href{http://xxx.lanl.gov/abs/1207.5192}{{\tt
  arXiv:1207.5192}}].

\bibitem{Susskind:2012uw}
L.~Susskind, {\it {The Transfer of Entanglement: The Case for Firewalls}},
  \href{http://xxx.lanl.gov/abs/1210.2098}{{\tt arXiv:1210.2098}}.

\bibitem{Papadodimas:2012aq}
K.~Papadodimas and S.~Raju, {\it {An Infalling Observer in AdS/CFT}},  {\em
  JHEP} {\bf 1310} (2013) 212, [\href{http://xxx.lanl.gov/abs/1211.6767}{{\tt
  arXiv:1211.6767}}].

\bibitem{Verlinde:2012cy}
E.~Verlinde and H.~Verlinde, {\it {Black Hole Entanglement and Quantum Error
  Correction}},  {\em JHEP} {\bf 1310} (2013) 107,
  [\href{http://xxx.lanl.gov/abs/1211.6913}{{\tt arXiv:1211.6913}}].

\bibitem{Maldacena:2013xja}
J.~Maldacena and L.~Susskind, {\it {Cool horizons for entangled black holes}},
  \href{http://xxx.lanl.gov/abs/1306.0533}{{\tt arXiv:1306.0533}}.

\bibitem{Susskind:2013lpa}
L.~Susskind, {\it {New Concepts for Old Black Holes}},
  \href{http://xxx.lanl.gov/abs/1311.3335}{{\tt arXiv:1311.3335}}.

\bibitem{Susskind:2014rva}
L.~Susskind, {\it {Computational Complexity and Black Hole Horizons}},
  \href{http://xxx.lanl.gov/abs/1402.5674}{{\tt arXiv:1402.5674}}.

\bibitem{Susskind:2014ira}
L.~Susskind, {\it {Addendum to Computational Complexity and Black Hole
  Horizons}},  \href{http://xxx.lanl.gov/abs/1403.5695}{{\tt arXiv:1403.5695}}.

\bibitem{Bousso:2013ifa}
R.~Bousso, {\it {Frozen Vacuum}},  {\em Phys.Rev.Lett.} {\bf 112} (2014)
  041102, [\href{http://xxx.lanl.gov/abs/1308.3697}{{\tt arXiv:1308.3697}}].

\bibitem{Banks:1998dd}
T.~Banks, M.~R. Douglas, G.~T. Horowitz, and E.~J. Martinec, {\it {AdS dynamics
  from conformal field theory}},
  \href{http://xxx.lanl.gov/abs/hep-th/9808016}{{\tt hep-th/9808016}}.

\bibitem{Hamilton:2006az}
A.~Hamilton, D.~N. Kabat, G.~Lifschytz, and D.~A. Lowe, {\it {Holographic
  representation of local bulk operators}},  {\em Phys.Rev.} {\bf D74} (2006)
  066009, [\href{http://xxx.lanl.gov/abs/hep-th/0606141}{{\tt
  hep-th/0606141}}].

\bibitem{Kabat:2011rz}
D.~Kabat, G.~Lifschytz, and D.~A. Lowe, {\it {Constructing local bulk
  observables in interacting AdS/CFT}},  {\em Phys.Rev.} {\bf D83} (2011)
  106009, [\href{http://xxx.lanl.gov/abs/1102.2910}{{\tt arXiv:1102.2910}}].

\bibitem{Heemskerk:2012mn}
I.~Heemskerk, D.~Marolf, J.~Polchinski, and J.~Sully, {\it {Bulk and
  Transhorizon Measurements in AdS/CFT}},  {\em JHEP} {\bf 1210} (2012) 165,
  [\href{http://xxx.lanl.gov/abs/1201.3664}{{\tt arXiv:1201.3664}}].

\bibitem{Freivogel:2004rd}
B.~Freivogel and L.~Susskind, {\it {A Framework for the landscape}},  {\em
  Phys.Rev.} {\bf D70} (2004) 126007,
  [\href{http://xxx.lanl.gov/abs/hep-th/0408133}{{\tt hep-th/0408133}}].

\bibitem{Horowitz:2009wm}
G.~Horowitz, A.~Lawrence, and E.~Silverstein, {\it {Insightful D-branes}},
  {\em JHEP} {\bf 0907} (2009) 057,
  [\href{http://xxx.lanl.gov/abs/0904.3922}{{\tt arXiv:0904.3922}}].

\bibitem{Heemskerk:2012np}
I.~Heemskerk, {\it {Construction of Bulk Fields with Gauge Redundancy}},  {\em
  JHEP} {\bf 1209} (2012) 106, [\href{http://xxx.lanl.gov/abs/1201.3666}{{\tt
  arXiv:1201.3666}}].

\bibitem{Aharony:2008ug}
O.~Aharony, O.~Bergman, D.~L. Jafferis, and J.~Maldacena, {\it {N=6
  superconformal Chern-Simons-matter theories, M2-branes and their gravity
  duals}},  {\em JHEP} {\bf 0810} (2008) 091,
  [\href{http://xxx.lanl.gov/abs/0806.1218}{{\tt arXiv:0806.1218}}].

\bibitem{Harlow:2011ke}
D.~Harlow and D.~Stanford, {\it {Operator Dictionaries and Wave Functions in
  AdS/CFT and dS/CFT}},  \href{http://xxx.lanl.gov/abs/1104.2621}{{\tt
  arXiv:1104.2621}}.

\bibitem{Leichenauer:2013kaa}
S.~Leichenauer and V.~Rosenhaus, {\it {AdS black holes, the bulk-boundary
  dictionary, and smearing functions}},  {\em Phys.Rev.} {\bf D88} (2013)
  026003, [\href{http://xxx.lanl.gov/abs/1304.6821}{{\tt arXiv:1304.6821}}].

\bibitem{Donnelly:2011hn}
W.~Donnelly, {\it {Decomposition of entanglement entropy in lattice gauge
  theory}},  {\em Phys.Rev.} {\bf D85} (2012) 085004,
  [\href{http://xxx.lanl.gov/abs/1109.0036}{{\tt arXiv:1109.0036}}].

\bibitem{Donnelly:2012st}
W.~Donnelly and A.~C. Wall, {\it {Do gauge fields really contribute negatively
  to black hole entropy?}},  {\em Phys.Rev.} {\bf D86} (2012) 064042,
  [\href{http://xxx.lanl.gov/abs/1206.5831}{{\tt arXiv:1206.5831}}].

\bibitem{Casini:2013rba}
H.~Casini, M.~Huerta, and J.~A. Rosabal, {\it {Remarks on entanglement entropy
  for gauge fields}},  \href{http://xxx.lanl.gov/abs/1312.1183}{{\tt
  arXiv:1312.1183}}.

\bibitem{Radicevic:2014kqa}
D.~Radicevic, {\it {Notes on Entanglement in Abelian Gauge Theories}},
  \href{http://xxx.lanl.gov/abs/1404.1391}{{\tt arXiv:1404.1391}}.

\bibitem{Streater:1989vi}
R.~Streater and A.~Wightman, {\em {PCT, spin and statistics, and all that}}.
\newblock 1989.

\bibitem{Hartle:1976tp}
J.~Hartle and S.~Hawking, {\it {Path Integral Derivation of Black Hole
  Radiance}},  {\em Phys.Rev.} {\bf D13} (1976) 2188--2203.

\bibitem{Israel:1976ur}
W.~Israel, {\it {Thermo field dynamics of black holes}},  {\em Phys.Lett.} {\bf
  A57} (1976) 107--110.

\bibitem{Maldacena:2001kr}
J.~M. Maldacena, {\it {Eternal black holes in anti-de Sitter}},  {\em JHEP}
  {\bf 0304} (2003) 021, [\href{http://xxx.lanl.gov/abs/hep-th/0106112}{{\tt
  hep-th/0106112}}].

\bibitem{Hayden:2007cs}
P.~Hayden and J.~Preskill, {\it {Black holes as mirrors: Quantum information in
  random subsystems}},  {\em JHEP} {\bf 0709} (2007) 120,
  [\href{http://xxx.lanl.gov/abs/0708.4025}{{\tt arXiv:0708.4025}}].

\bibitem{Srednicki:1995pt}
M.~Srednicki, {\it {Thermal fluctuations in quantized chaotic systems}},  {\em
  J.Phys.} {\bf A29} (1996) L75--L79,
  [\href{http://xxx.lanl.gov/abs/chao-dyn/9511001}{{\tt chao-dyn/9511001}}].

\bibitem{lloydthesis}
S.~Lloyd, {\em Black Holes, Demons, and the Loss of Coherence: How complex
  systems get information, and what they do with it. (PhD Thesis)}.
\newblock http://meche.mit.edu/people/?id=55, 1988.

\bibitem{Shenker:1990uf}
S.~H. Shenker, {\it {The Strength of nonperturbative effects in string
  theory}}, .

\bibitem{Verlinde:2013qya}
E.~Verlinde and H.~Verlinde, {\it {Behind the Horizon in AdS/CFT}},
  \href{http://xxx.lanl.gov/abs/1311.1137}{{\tt arXiv:1311.1137}}.

\bibitem{Shenker:2013pqa}
S.~H. Shenker and D.~Stanford, {\it {Black holes and the butterfly effect}},
  \href{http://xxx.lanl.gov/abs/1306.0622}{{\tt arXiv:1306.0622}}.

\bibitem{Shenker:2013yza}
S.~H. Shenker and D.~Stanford, {\it {Multiple Shocks}},
  \href{http://xxx.lanl.gov/abs/1312.3296}{{\tt arXiv:1312.3296}}.

\bibitem{Rocha:2008fe}
J.~V. Rocha, {\it {Evaporation of large black holes in AdS: Coupling to the
  evaporon}},  {\em JHEP} {\bf 0808} (2008) 075,
  [\href{http://xxx.lanl.gov/abs/0804.0055}{{\tt arXiv:0804.0055}}].

\bibitem{Harlow:2013tf}
D.~Harlow and P.~Hayden, {\it {Quantum Computation vs. Firewalls}},  {\em JHEP}
  {\bf 1306} (2013) 085, [\href{http://xxx.lanl.gov/abs/1301.4504}{{\tt
  arXiv:1301.4504}}].

\bibitem{harlow}
D.~Harlow, {\em Unpublished}.
\newblock 2013.

\bibitem{aaronson}
S.~Aaronson, {\it To appear}, .

\bibitem{preskillnotes}
J.~Preskill, {\em Lecture Notes on Quantum Computation}.
\newblock http://www.theory.caltech.edu/people/preskill/ph229/, 1998.

\bibitem{Horowitz:2003he}
G.~T. Horowitz and J.~M. Maldacena, {\it {The Black hole final state}},  {\em
  JHEP} {\bf 0402} (2004) 008,
  [\href{http://xxx.lanl.gov/abs/hep-th/0310281}{{\tt hep-th/0310281}}].

\bibitem{Gottesman:2003up}
D.~Gottesman and J.~Preskill, {\it {Comment on `The Black hole final state'}},
  {\em JHEP} {\bf 0403} (2004) 026,
  [\href{http://xxx.lanl.gov/abs/hep-th/0311269}{{\tt hep-th/0311269}}].

\bibitem{Lloyd:2013bza}
S.~Lloyd and J.~Preskill, {\it {Unitarity of black hole evaporation in
  final-state projection models}},
  \href{http://xxx.lanl.gov/abs/1308.4209}{{\tt arXiv:1308.4209}}.

\bibitem{Bousso:2013uka}
R.~Bousso and D.~Stanford, {\it {Measurements without Probabilities in the
  Final State Proposal}},  {\em Phys.Rev.} {\bf D89} (2014) 044038,
  [\href{http://xxx.lanl.gov/abs/1310.7457}{{\tt arXiv:1310.7457}}].

\bibitem{Mathur:2013gua}
S.~D. Mathur and D.~Turton, {\it {The flaw in the firewall argument}},  {\em
  Nucl.Phys.} {\bf B884} (2014) 566,
  [\href{http://xxx.lanl.gov/abs/1306.5488}{{\tt arXiv:1306.5488}}].

\bibitem{Harlow:2010my}
D.~Harlow and L.~Susskind, {\it {Crunches, Hats, and a Conjecture}},
  \href{http://xxx.lanl.gov/abs/1012.5302}{{\tt arXiv:1012.5302}}.

\bibitem{Bousso:2011up}
R.~Bousso and L.~Susskind, {\it {The Multiverse Interpretation of Quantum
  Mechanics}},  {\em Phys.Rev.} {\bf D85} (2012) 045007,
  [\href{http://xxx.lanl.gov/abs/1105.3796}{{\tt arXiv:1105.3796}}].

\bibitem{Bekenstein:1980jp}
J.~D. Bekenstein, {\it {A Universal Upper Bound on the Entropy to Energy Ratio
  for Bounded Systems}},  {\em Phys.Rev.} {\bf D23} (1981) 287.

\bibitem{Casini:2008cr}
H.~Casini, {\it {Relative entropy and the Bekenstein bound}},  {\em
  Class.Quant.Grav.} {\bf 25} (2008) 205021,
  [\href{http://xxx.lanl.gov/abs/0804.2182}{{\tt arXiv:0804.2182}}].

\bibitem{Wald:1993nt}
R.~M. Wald, {\it {Black hole entropy is the Noether charge}},  {\em Phys.Rev.}
  {\bf D48} (1993) 3427--3431,
  [\href{http://xxx.lanl.gov/abs/gr-qc/9307038}{{\tt gr-qc/9307038}}].

\bibitem{Iyer:1994ys}
V.~Iyer and R.~M. Wald, {\it {Some properties of Noether charge and a proposal
  for dynamical black hole entropy}},  {\em Phys.Rev.} {\bf D50} (1994)
  846--864, [\href{http://xxx.lanl.gov/abs/gr-qc/9403028}{{\tt
  gr-qc/9403028}}].

\bibitem{headrick}
M.~Headrick, V.~Hubeny, A.~Lawrence, and M.~Rangamani, {\it {Causality and
  holographic entanglement entropy (in preperation)}}, .

\bibitem{harlow2}
D.~Harlow, {\it {Complementarity, not Firewalls}},
  \href{http://xxx.lanl.gov/abs/1207.6243}{{\tt arXiv:1207.6243}}.

\end{thebibliography}\endgroup
\end{document}